\pdfoutput=1
\documentclass[traditabstract,a4paper]{aa}
\usepackage{graphicx}
\usepackage{txfonts}
\usepackage[authoryear]{natbib}
\newcommand{\nh}{N$_{\rm H}$}
\newcommand{\Halp}{H${\alpha}$}

\newcommand{\ergscm}{erg\,cm$^{-2}$\,s$^{-1}$}
\newcommand{\ergcms}{erg\,cm$^{-2}$\,s$^{-1}$}
\newcommand{\Fx}{$\rm F_{\rm X}$}
\newcommand{\Lx}{$\rm L_{\rm X}$}
\newcommand{\ergs}{erg s$^{-1}$}

\newcommand{\Teff}{$T_{\rm eff}$\,}

\newcommand{\logg}{$\log g$\,}	
\begin{document}
%
   \title{Cross-correlation of the 2XMMi catalogue with Data Release 7 of the Sloan Digital Sky Survey\thanks{The corresponding fits file can be downloaded from the XCat-DB home page (http://xcatdb.u-strasbg.fr/). The file also contains line information for all SDSS spectroscopic entries matching a 2XMM source. Results from the cross-correlation with the 2XMM DR3 are also available at the same location.}}


   \author{F.-X. Pineau\inst{1}
          \and
	  C. Motch\inst{1}
	  \and 
	  F. Carrera\inst{2}
	  \and 
	  R. Della Ceca\inst{3}
	  \and 
	  S. Derri\`ere\inst{1}
	  \and 
	  L. Michel\inst{1}
	  \and 
	  A. Schwope\inst{4}
	  \and
	  M.G. Watson \inst{5}
          }

   \institute{
         CNRS, Universit\'e de Strasbourg, Observatoire Astronomique, 11 rue de l'Universit\'e, 67000 Strasbourg, France\\
              \email{francois-xavier.pineau \emph{at} astro.unistra.fr} 
         \and
	 Instituto de F\`isica de Cantabria (CSIC-UC), 39005 Santander, Spain
	 \and
	 INAF - Osservatorio Astronomico di Brera, via Brera 28, 20121 Milan, Italy
	 \and
	 Astrophysikalisches Institut Potsdam, An der Sternwarte 16, 14482 Potsdam, Germany
	 \and
	 Department of Physics and Astronomy, University of Leicester, LE1 7RH, UK
           }

   \date{Received, accepted }

  \abstract{The Survey Science Centre of the XMM-Newton satellite released the first incremental version of the 2XMM catalogue in August 2008 . Containing more than 220,000 X-ray sources, the 2XMMi was at that time the largest catalogue of X-ray sources ever published and thus constitutes an unprecedented resource for studying the high-energy properties of various classes of X-ray emitters such as AGN and stars. Thanks to the high throughput of the EPIC cameras on board XMM-Newton accurate positions, fluxes, and hardness ratios are available for a substantial fraction of the X-ray detections. The advent of the 7$^{th}$ release of the Sloan Digital Sky Survey offers the opportunity to cross-match two major surveys and extend the spectral energy distribution of many 2XMMi sources towards the optical bands. This implies building extensive homogeneous samples with a statistically controlled rate of spurious matches and completeness. We here present a cross-matching algorithm based on the classical likelihood ratio estimator. The method developed has the advantage of providing true probabilities of identifications without resorting to heavy Monte-Carlo simulations. Over 30,000 2XMMi sources have SDSS counterparts with individual probabilities of identification higher than 90\%. At this threshold, the sample has only 2\% spurious matches and contains 77\% of all expected SDSS identifications. Using spectroscopic identifications from the SDSS DR7 catalogue supplemented by extraction from other catalogues, we build an identified sample from which the way the various classes of X-ray emitters gather in the multi dimensional parameter space can be analysed and later used to design a source classification scheme. We illustrate the interest of this clean source sample by investigating two scientific use cases. In the first example we show how these multi-wavelength data can be used to search for new QSO2s. Although no specific range of observed properties allows us to efficiently identify Compton Thick QSO2s, we show that the prospects are much better for Compton Thin AGN2 and discuss several possible multi-parameter selection strategies. In a second example, we confirm the hardening of the mean X-ray spectrum with increasing X-ray luminosity on a sample of over 500 X-ray active stars and reveal that on average X-ray active M stars display bluer $g-r$ colour indexes than less active ones. Although this catalogue of 2XMM-SDSS sources cannot be used directly for statistical studies, it nevertheless represents an excellent starting point to select well defined samples of X-ray-emitting objects.}  
  

   \maketitle

\section{Introduction}

The growing collecting area and sensitivity of modern astronomical detectors combined with the  increasing storage and processing capabilities offered by current computer facilities has made possible the gathering on comparatively short time scales of very large sky surveys that were beyond reach only a few years ago. Most parts of the electromagnetic spectrum benefit from this evolution. Among recently completed or ongoing projects are the Two Micron All Sky Survey (2MASS) \citep{2003tmc..book.....C} and the Sloan Digital Sky Survey \citep{2008ApJS..175..297A} for instance. Space-borne missions currently in operation such as the Spitzer Space Telescope \citep{2004ApJS..154....1W} observing in the infra-red or the Chandra \citep{2000SPIE.4012....2W} and XMM-Newton \citep{jansen2001} X-ray observatories are collecting at a high rate a wealth of measurements on an unprecedented number of objects in their energy range. In the relatively near future, ground-based automated very large telescopes such as pan-STARRS \citep{wang2010} or such as the Large Synoptic Survey Telescope \citep{ 2002SPIE.4836...10T} will collect detailed photometric information on a breathtaking number of faint galaxies.  

Merging measurements arising from several instruments allows us to build spectral energy distributions in a range of wavelengths extending over a large part of the electromagnetic spectrum. The recent availability of wide angle surveys with high detection sensitivities allows us to measure with comparable accuracies and in several scientifically important wavelength ranges the spectral energy density of the main classes of X-ray emitting astrophysical sources. Building large homogeneous samples provides valuable insight on the emission mechanisms and evolutionary processes and may allow the detection of rare objects or outliers, which would be otherwise hard to unveil in smaller samples. In this respect, a good estimate of the true rate of false cross-identification is important to assess the relevance of any group of outliers.

However, the gathering of large groups of sources with well characterised multi-wavelength properties first requires a proper handling of the cross-matching process between two or more catalogues. Although spatial resolution at high-energy steadily increased during the last years and may go on improving in the future, source density also grows as a result of the improved sensitivity, and the risk of confusion between unrelated objects detected at different wavelengths does not necessarily vanish. The confusion problem can be particularly arduous when comparing catalogues with very different spatial resolutions and densities, a problem often encountered in the identification process of high-energy sources which in several cases lack the superb spatial resolution affordable for instance in the optical domain, see e.g. \citet[][]{2000ApJS..131..335R} for the identification of ROSAT sources and \citet[][]{luo2010} for a recent example involving multi-wavelength catalogues with different depths and angular resolutions. 

The XMM-Newton satellite \citep{jansen2001} was launched by the European Space Agency late in 1999. XMM-Newton is currently the X-ray (0.2-12\,keV) telescope in operation with the largest effective area. Three co-aligned telescopes feed two EPIC MOS \citep{turner2001} and one EPIC pn \citep{struder2001} cameras. Two reflection grating arrays deviate about half of the X-ray photons from the EPIC MOS camera towards two Reflection Grating Spectrometers \citep[RGS;][]{denherder2001}. An Optical Monitor \citep[OM;][]{mason2001}, providing UV and optical images of a fraction of the field of view covered by the EPIC cameras down to the 21$^{th}$ magnitude, complements the X-ray instrumentation. One of the remarkable properties offered by the X-ray telescopes on-board XMM-Newton is to provide a large field of view of 30\arcmin\ diameter with a weakly degraded image point-spread function and low vignetting even at large off-axis angles. Accordingly, a large number of sources may be serendipitously discovered around the main target of the observation, which builds up to make an X-ray survey with an unprecedented combination of sensitivity and area covered. Starting from the beginning of the project, ESA recognised the high scientific interest of exploiting the XMM-Newton survey and appointed the present Survey Science Centre (SSC) on a competitive basis. Lead by the University of Leicester, the SSC is a consortium of ten European institutes conducting its activity on behalf of ESA. The SSC responsibilities have been presented in \cite{watson2001}. One of the most demanding tasks given to the consortium is the compilation of a catalogue of all sources serendipitously  discovered in the field of view of the X-ray instruments and of their characterisation and identification at least in a statistical way. 

Several spectroscopic identification campaigns and multi-wavelength studies have been recently performed by the SSC on samples of thousands of EPIC sources using follow-up observations at 4-m and 8-m class telescopes. The availabilities of the recently published SDSS Data Release 7 (DR7) and of the incremental version of the 2XMM catalogue (2XMMi) offer a unique opportunity to extend the identification work to a much more extended sky area. With its spectroscopic and photometric limiting magnitude about 2 magnitudes brighter than that typically reached for the SSC source samples, SDSS identifications of XMM-Newton sources conveniently expand the identified sample towards brighter magnitudes and at the same time provide access to a rich group of accurately quantified photometric and spectroscopic data. 

As part of its scientific activities, the Survey Science Centre of the XMM-Newton satellite has developed a specific cross-correlation algorithm yielding actual probabilities of identification based on positional coincidence and applied this algorithm to the cross-identification of the 2XMMi and SDSS DR7 catalogues, thus creating one of the largest set of optically identified X-ray sources available so far. The result of the cross-correlation is made available as a separate fits file and is also available through the XCat-DB\footnote{http://xcatdb.u-strasbg.fr/} \citep{motch2007aspc,michel2009}.

The first sections of this paper present the details of the algorithm used to identify 2XMMi X-ray sources with SDSS DR7 optical objects. We apply the commonly used likelihood ratio to quantify the chance that a SDSS object is the counterpart of the X-ray source. Identification probabilities are computed with an original method that does not rely on Monte Carlo simulations and thus offers a better efficiency when cross-correlating large sets of data. We then describe the range of optical and X-ray parameters occupied by the main astrophysical classes of X-ray emitters and show how source classification could be achieved on this basis. In the last part of this paper, we investigate two example science cases, the search for new QSO2s, and the study of the properties of the X-ray active late-type star population.   

\section{Description of the cross-correlated catalogues}
        
\subsection{2XMMi catalogue}

The incremental Second XMM-Newton Serendipitous Source Catalogue (2XMMi) is an extended version of the 2XMM Catalogue \citep{watson2009}. It has been built from 4117 individual pointed observations performed by the XMM-Newton Observatory and contains 289\,083 heterogeneous detections for a total of 221\,012 unique X-ray sources. The catalogue covers $\sim 1\%$ of the sky over a large range of Galactic latitudes and longitudes. Owing to the wide range of exposure times, the area covered  sensitively depends on limiting flux and energy range \citep[see Fig. 8 in][]{watson2009}. A 90\% complete relative sky coverage is reached at \Fx\,=\,1 and 9 $\times$ 10$^{-14}$\ergscm\ in the 0.5-2.0\,keV and 2.0-12.0\,keV bands respectively. The EPIC cameras encompass a field (FOV) of $\sim 30'$ diameter and are sensitive in the energy range of $\sim 0.2$ -- $12$ keV. Source positions have a typical accuracy of $\sim 2''$. In this paper, we limit our analysis to point-like sources with a positional error smaller or equal to $5''$. A source is defined as point-like if its extent maximum likelihood parameter (\emph{ep\_ext\_ml}) is $\le 4$. The resulting 2XMMi source sample consists of 264,361 detections and 200,067 unique 2XMMi sources.
  
\subsection{SDSS Data Release 7}

The Seventh Data Release of the Sloan Digital Sky Survey \citep{abazajian2009}, covers $11663$\,deg$^{2}$ , mostly in the northern Galactic cap. A total of 357 million objects have 5 band photometry, among which 1.6 million galaxies, quasars, and stars were spectroscopically observed. Most of the $\sim$ 2000\,deg$^{2}$ increment over data release 6 are located at low galactic latitude. Astrometric errors are $< 0.1''$ $\mathrm{ rms}$. At the 3\% error level, the catalogue reaches magnitude limits in the range of $20.5$ to $22.2$ in the five photometric bands -- u, g, r, i and z --.
In this paper we only consider the so-called primary sources of the SDSS DR7 Photometric Catalogue as available from the VizieR data server. Primary sources are the ``main'' detection of an object and have the best defined set of parameters. For most scientific applications, the primary detections are the only ones needed. Source lists have been extracted using the VO ConeSearch protocol. The central point of each query is the centre of the FOV of the XMM-Newton observation considered and the search radius is the distance from the centre to the farthest X-ray source, to which we add $3'$ for completeness.

\section{Counterpart identification procedure}

We discuss in (\ref{sec:optcandselect}) how we select optical candidates, taking into account arbitrary error ellipses on the source's spherical coordinates. We compute a likelihood ratio ($LR$) for each target-candidate pair (\ref{sec:lr}). This $LR$ involves a measure of the local density using a kernel smoothing method (Appendix \ref{part:density_estimation}). Estimating the true $LR$ distribution for spurious associations (\ref{sec:reliability}) then allows us to compute for each target-candidate pair the probability of association only based on positional coincidence (\ref{sec:proba_of_id}). 

\subsection{Selection of optical candidates \label{sec:optcandselect}}

\subsubsection{Selection criterion}

We consider a target X-ray source and a candidate optical source with $\alpha_X, \delta_X$ the equatorial coordinates of the X-ray source; $\sigma_{\alpha_X}$, $\sigma_{\delta_X}$ and $\rho_{X_{\alpha,\delta}}$ the error on $\alpha_X \cos \delta_X$ and on $\delta_X$ and the correlation between $\sigma_{\alpha_X}$ and $\sigma_{\delta_X}$ respectively; 
$\alpha_o \mbox{, } \delta_o$ the equatorial coordinates of an optical source $\sigma_{\alpha_o}$, $\sigma_{\delta_o}$ and $\rho_{o_{\alpha,\delta}}$ the error on $\alpha_o \cos \delta_o$ and on $\delta_o$ and the correlation between $\sigma_{\alpha_o}$ and $\sigma_{\delta_o}$, respectively. 

\begin{figure}
    \centering
    \includegraphics[width=0.48\textwidth]{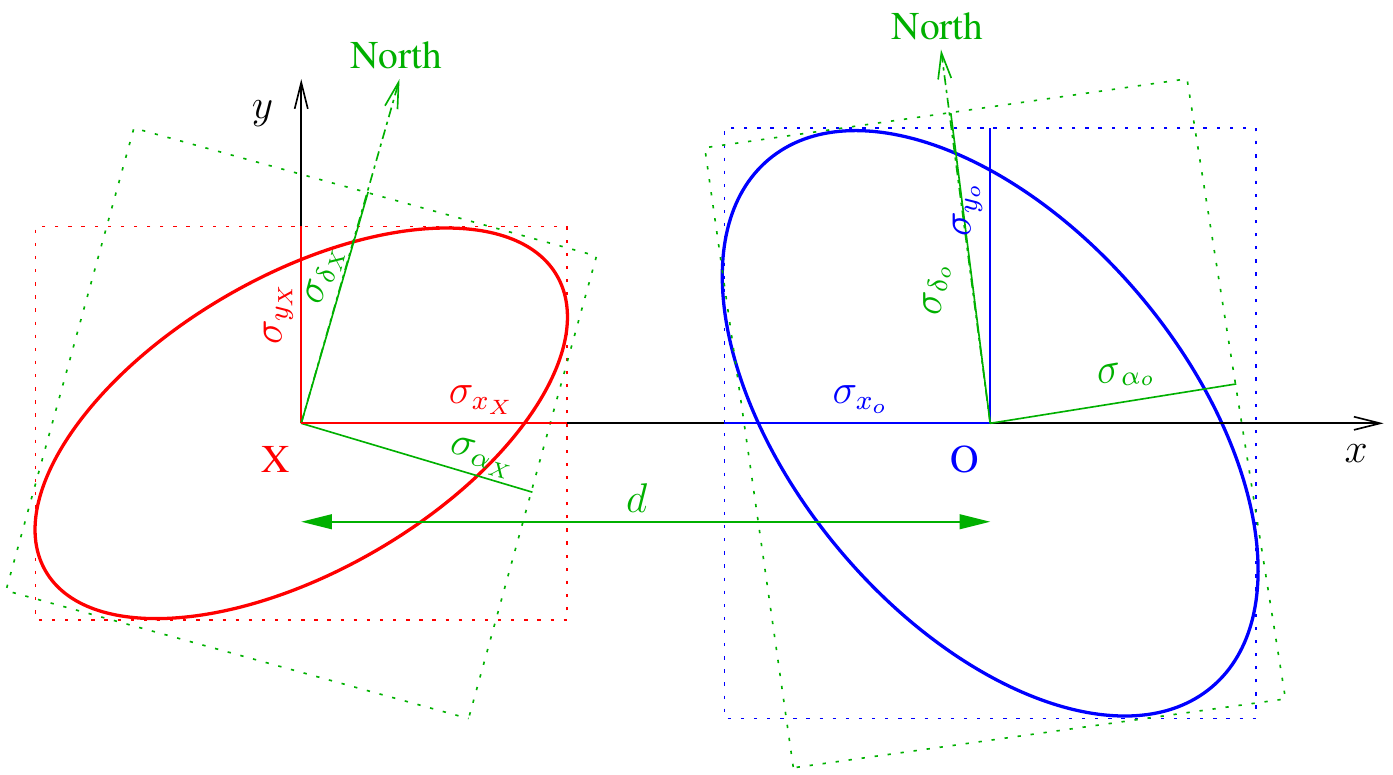}
    \caption{Chosen projection plane: the $xy$ frame is centred on the X-ray source position $X$;
             the $x$-axis is the direction towards the optical candidate, located at point $O$.
             $d$ is the angular distance between the two sources.
             This frame is useful at high declinations when we cannot consider 
             the meridians any longer -- the directions of the north pole in $X$ and in $O$ -- to be parallel. It allows
             us to deal naturally with the poles.
    }
    \label{fig:ellipses}
\end{figure}

As everybody implicitly does -- except \citet{2008ApJ...679..301B} --, we convert the spherical problem into a plane one and positional errors are interpreted as usual 2D Gaussians. We have chosen a projection on a 2D plane with a frame centred on the position of the X-ray source and having for $x$-axis the direction of the optical candidate (Fig. \ref{fig:ellipses}). Errors on positions become Gaussians: $\mathcal{N}_X(x,y ; \sigma_{x_X}^2 , \sigma_{y_X}^2 , \rho_X \sigma_{x_X} \sigma_{y_X})$ and $\mathcal{N}_o(x-d,y;\sigma_{x_o}^2,\sigma_{y_o}^2,\rho_o \sigma_{x_o} \sigma_{y_o})$ with $d$ the angular distance between the X-ray and the optical source. As suggested by \citet{1984S&T....68R.158S}, $d$ is computed using the Haversine function. The transformation of the ellipses in the new reference frame is described in Appendix \ref{apA}.

The density of probability that the two sources are at the same location, and thus are the same object, is given by
the convolution product of these two distributions. It leads to a new Gaussian:
\begin{equation}
    P(x,y) = \mathcal{N}_c(x,y; \sigma_{x_c},\sigma_{y_c},\rho_c\sigma_{x_c}\sigma_{y_c}),
    \label{eq:Nc}
\end{equation}
With $\sigma_{x_c}^2 = \sigma_{x_X}^2 + \sigma_{x_o}^2$, $\sigma_{y_c}^2=\sigma_{y_X}^2 + \sigma_{y_o}^2$ and $\rho_c\sigma_{x_c}\sigma_{y_c}=\rho_X \sigma_{x_X} \sigma_{y_X} + \rho_o \sigma_{x_o} \sigma_{y_o}$.

If the optical source is the counterpart of the X-ray source, it falls 
with a probability $\gamma$ inside the ellipse defined by the equation
\begin{equation}
    \Big (
        \begin{array}{c}
            x\\
            y\\
        \end{array}
    \Big )^t
    \Big (
        \begin{array}{cc}
            \sigma_{x_c}^2                & \rho_c\sigma_{x_c}\sigma_{y_c}\\
            \rho_c\sigma_{x_c}\sigma_{y_c} & \sigma_{y_c}^2\\
        \end{array}
    \Big )^{-1}
    \Big (
        \begin{array}{c}
            x\\
            y\\
        \end{array}
    \Big )
    = k_{\gamma}^2 \ .
\end{equation}
The completeness we have chosen is a $3\sigma$ criterion, often used as a compromise between the total number of associations
and the number of counterparts missed ($0.3\%$).
This completeness, $\gamma=99.7\%$, leads in 2D to $k_{\gamma}=3.43935$.
In the frame we have chosen, the coordinates of the optical source are $x=d$ and $y=0$.
The selection criterion we adopt will retain all candidates satisfying
\begin{equation}
    \frac{d}{\sigma_{x_c}\sqrt{1-(\rho_c\sigma_{x_c}\sigma_{y_c})^2}} \leq k_\gamma \ .
    \label{eq:criterion}
\end{equation}

We make the additional following hypotheses. First, we neglect any systematic offset between the positions of the two catalogues. The 2XMMi catalogue as a whole is free of any systematic positional offset in a direction of the sky. This has been checked by cross-correlating the 2XMM catalogue with the SDSS DR5 Quasar catalogue \citep{watson2009}. For a large number of cases (74\% at $|b| > 20\degr$), it was possible to correct the astrometry by cross-correlating field X-ray sources with USNO B1.0 entries. When no reliable astrometric correction could be found, increasing the applied systematic error from 0.35\arcsec to 1.0\arcsec accounts for the possible remaining coordinate offset and rotation affecting all the sources detected in a given observation \citep{watson2009}. 
Second, we assume that all positions and associated errors have been computed at the same epoch and therefore corrected for proper motions. 

\subsubsection{Application to XMM-SDSS DR7 data}

The 2XMMi catalogue provides a circular error on position ($radec\_err$) and a systematic error ($syserr$) for each source .
The error on positions of each X-ray source is the quadratic sum of these two values: 
\begin{equation}
    \sigma_{\alpha_X} = \sigma_{\delta_X} = \sqrt{radec\_err^2 + syserr^2} \ .
\end{equation}
Because it is symmetric, we have $\rho_{X_{\alpha,\delta}}=0$, $\rho_X=0$ and thus $\rho_c\sigma_{x_c}\sigma_{y_c}$ is directly
equal to $\rho_o \sigma_{x_o} \sigma_{y_o}$.

Positional errors are elliptical in the SDSS DR7 catalogue: $\sigma_{\alpha_o}=raErr$, $\sigma_{\delta_o}=decErr$
and $\rho_{o_{\alpha,\delta}}=raDecCorr$. The definitions of the different parameters are summarized in Table \ref{tab:astrom_params}.

\begin{table}
   \centering  
   \caption{Summary of the astrometric parameters for the 2XMM catalogue and for the SDSS DR7.} 
   \begin{tabular}{l|c|c}
         & 2XMMi & SDSS DR7 \\
	 \hline
         $\sigma_\alpha$ & $\sqrt{radec\_err^2 + syserr^2}$ & $raErr$ \\
	 $\sigma_\delta $  & $\sqrt{radec\_err^2 + syserr^2}$ & $decErr$ \\
	 $\rho_{\alpha,\delta}$  & 0 & $raDecCorr$\\
   \end{tabular}
   
  \label{tab:astrom_params}
\end{table}

\subsection{Likelihood ratio \label{sec:lr}}

We compute a likelihood ratio ($LR$) for each target-candidate pair meeting the criterion of Eq. \ref{eq:criterion}:
the probability of finding the optical counterpart at a normalised distance $r$  (see below) divided by the probability of having a spurious object at that distance.

The density of probability that the two sources are at the same location knowing $x$ and $y$ corresponds
to the density of probability of having the counterpart in $x$ and $y$, 
assuming that it is the same astrophysical object as the X-ray emitting one.
The Gaussian $\mathcal{N}_c$ (Eq. \ref{eq:Nc}) can be written in its canonical form $\frac{1}{2\pi\sigma_M\sigma_m}\exp{ -\frac{1}{2}(\frac{x_1^2}{\sigma_M^2} +\frac{y_1^2}{\sigma_m^2} ) }$.
Where $\sigma_M$ and $\sigma_m$  are the semi-major and semi-minor axis, in the eigenvector frame $(x_1,y_1)$,
given by the eigendecomposition of the variance-covariance matrix of $\mathcal{N}_c$,
\begin{equation}
  \sigma_{M,m} = \frac{1}{2}(\sigma_{x_c}^2+\sigma_{y_c}^2 \pm \sqrt{(\sigma_{x_c}^2-\sigma_{y_c}^2)^2+4(\rho\sigma_{x_c}\sigma_{y_c})^2}) \ .
\end{equation}
We change the scale and switch to polar coordinates, which leads to the dimensionless Rayleigh distribution:
\begin{equation}
    r =\sqrt{\frac{x_1^2}{\sigma_M^2} + \frac{y_1^2}{\sigma_m^2}}\ .
    \label{eq:r}
\end{equation}
Therefore, the new elementary surface becomes $\pi\sigma_M\sigma_m$, the surface of the $1\sigma$ (or $r=1$) ellipse.

The $LR$ we use is inspired by the one described in \citet{1977A&AS...28..211D}.
As \citet{1986MNRAS.223..279W}, we do not only consider the first candidate, but all sources satisfying 
Eq. \ref{eq:criterion}.
We thus replace the probability ``\emph{of finding the first confusing object at a distance lying between $r$ and $r+\mathrm{d}r$}''
by the one of finding \emph{a} confusing object between $r$ and $r+\mathrm{d}r$.

The probability of finding the optical counterpart (cp) at a distance lying between $r$ and $r+\mathrm{d}r$ is
\begin{equation}
    \mathrm{d} p(r|cp) = r e^{-\frac{1}{2}r^2} \mathrm{d}r \ .
    \label{eq:proba_cp}
\end{equation}
And the probability of finding a spurious object (spur) between $r$ an $r+\mathrm{d}r$ is given by the Poisson law:
\begin{equation}
    \mathrm{d} p(r|spur) = 2\lambda r \mathrm{d}r \ .
    \label{eq:proba_noise}
\end{equation}
We adopt the local surface density of sources at least as bright as $m_o$, the magnitude of the candidate. Because more sources are available in a same given area, the densities computed with this method are more local -- or more accurate -- than densities computed in arbitrary bins of magnitudes. It is equivalent to computing local densities using increasingly sensitive instruments. We detail in Appendix \ref{part:density_estimation} the method used to estimate local densities. 

The likelihood ratio is the ratio of the two probability densities (\ref{eq:proba_cp}) and (\ref{eq:proba_noise}):
\begin{equation}
    LR(r) = \frac{\mathrm{d} p(r|cp)}{\mathrm{d} p(r|spur)} = \frac{1}{2\lambda} e^{-\frac{1}{2}r^2} \ .
    \label{eq:lr}
\end{equation}

The formalism we apply here aims at providing probabilities of identification based on positional coincidences only. A Bayesian interpretation of the likelihood ratio method is described in Appendix \ref{LRBayesian}.
We do not use other information on sources such as the spectral energy distribution.
Hence, we do not add an extra term $q(m)$ to the $LR$ as is done for example in \citet{1986MNRAS.223..279W}, \citet{1992MNRAS.259..413S} and \citet{2007ApJS..172..353B}. The quantity $q(m)$ corresponds to the probability of having among the real counterparts a source of 
magnitude $m$, or in a bin $\Delta m$ around $m$ (see formula (\ref{eq:lrextended}) of the appendix). 
In this case, $q(m)$ should be local, but then becomes hard to estimate. In general the estimate of $q(m)$
is plagued with considerable errors which, besides the error on the local density estimation, dramatically affect the error on $LR$. We will see in Sect. \ref{sec:reliability} that the $q(m)$ factor is somehow taken into account in our reliability function.

\subsection{Computing reliabilities\label{sec:reliability}}

Although we use a different $LR$ definition, a different estimator of the rate of spurious associations and a different function
to fit the reliability histogram, we more or less follow the work presented in part 3 of \citet{2005AJ....130.2019O}.
The method originates in \citet{2000ApJS..131..335R}.

We define the reliability of an association in a given bin of $LR$ as
\begin{equation}
     R(LR) = \frac{N_{real}(LR)}{N_{real}(LR)+N_{spur}(LR)} = \frac{N_{cand}(LR)-N_{spur}(LR)}{N_{cand}(LR)}\ ,
    \label{eq:R}
\end{equation}
where $N_{real}$ and $N_{spur}$ are the unknown number of candidates which are respectively real and spurious counterparts 
in a given bin of $LR$; $N_{cand}$ is the number of candidates in a given bin of $LR$.

We therefore have to estimate $N_{spur}$. An often used method consists in correlating X-ray sources with artificial samples of optical sources. The generated samples have the same characteristics as the real sources: same density, same positional errors distribution, etc. Positions are randomly distributed. The sum of the results of these Monte-Carlo samples provides an estimate of the number of spurious associations as function of the distances, of the $LR$s, etc. This approach is used by \citet{2005AJ....130.2019O} among others. In \citet{2005A&A...432L..49S} the random sample consists in a list of \emph{``anti [...] sources''}, which are\emph{``mirrored in Galactic longitude and latitude''}.

We propose here a new method to estimate the number of spurious associations, not based on Monte-Carlo simulations, but instead directly computing their expected results. This scheme offers a better computing efficiency when cross-correlating huge sets of data.  The basic idea of estimating the surface of an association related to the total available area can be found in \citet{1998A&AS..129...87B}. The method is described in Appendix \ref{spurLR}.

In order to avoid computing too many local densities for estimating the rate of spurious associations, we divide the magnitude range into bins and associate all sources in the same magnitude bin with the mean value of their local density. The width of the bins depends on the magnitude accuracy of the catalogue. We then compute for all optical and X-ray sources the $\sigma_M \sigma_m$ factor, the $LR_{min}$ and $LR_{max}$. It is thus possible to compute the histogram of the expected number of spurious associations according to $LR$ values. To increase the computing efficiency, we can bin the $\sigma_M \sigma_m$ values. However, this approach involves another loss of accuracy for a meagre reduction of computing time.

\begin{figure}
\centering
\includegraphics[width=0.5\textwidth]{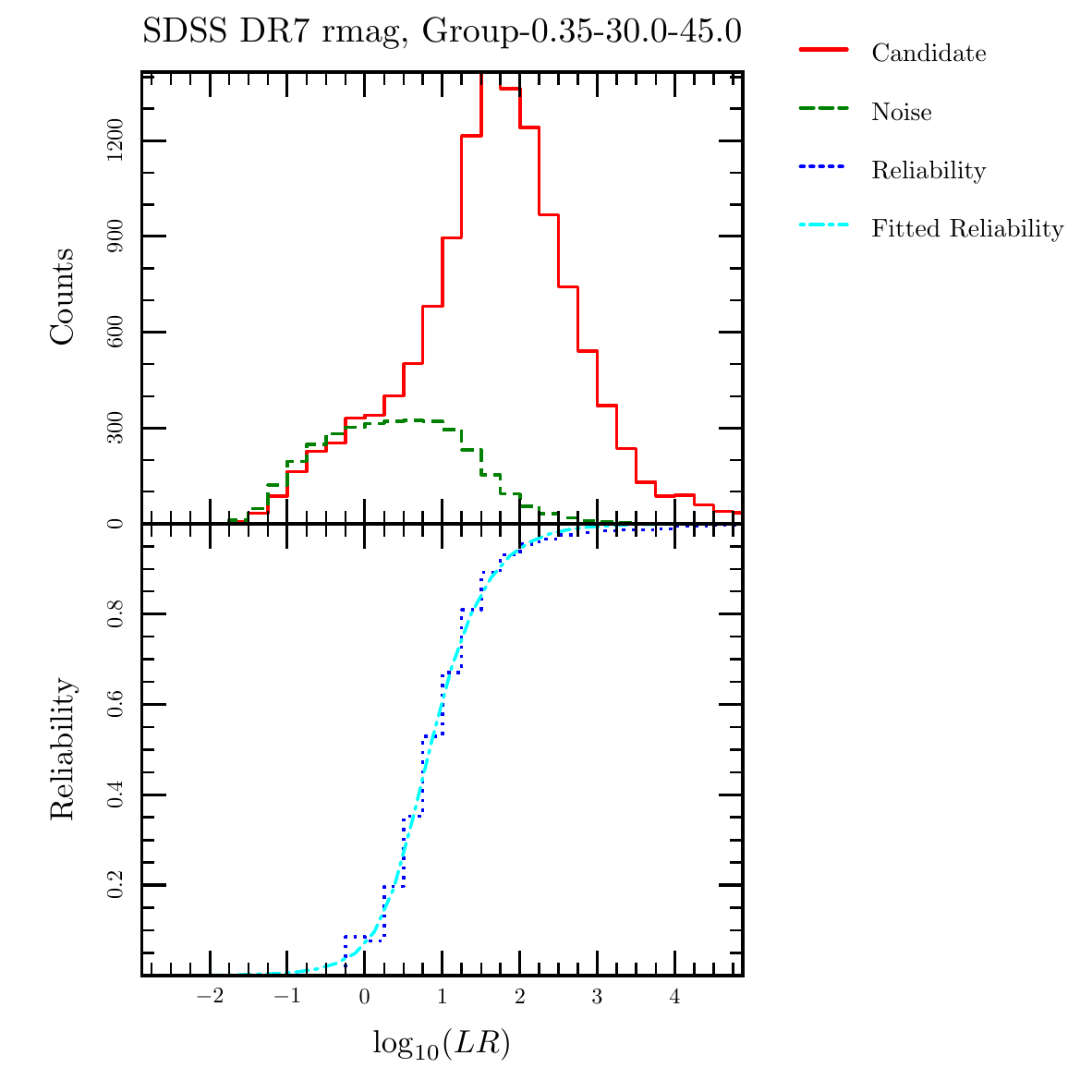}
\caption{Top: Histograms of the number of associations and of the estimate of the number of spurious associations by bin of $\log_{10}(LR)$. Bottom: Reliability histogram by bin of $\log_{10}(LR)$ and its fitted curve. In the example, $LR$s have been computed according to the SDSS DR7 $r$ magnitude for XMM sources with a systematic error of $0.35 \mbox{ } arcsec$ with a galactic latitude $30\degr<|b|<45\degr$.
}
\label{fig:histo_fit}
\end{figure}

As shown in Fig. \ref{fig:histo_fit}, the histograms used in the computation of the reliability are the number of candidates and the number of spurious associations grouped in bin of $\log_{10}{LR}$.

\subsubsection{Fitting the reliability function\\}

In order to estimate the number of spurious associations we take the relatively realistic example where the X-ray source has at most one candidate. The reliability of an association (not to be confused with the integrated reliability for all associations having a $LR \geq l $) is directly given by (see Eq. (\ref{eq:phcp}) of the appendix)
\begin{equation}
    R(LR)=p(id|r)=\frac{1}{1+\frac{p(spur)}{p(cp)}\frac{1}{LR}} \ .
    \label{eq:r_ideal}
\end{equation}
The term $p(cp)/p(spur)$ -- the probability that the optical source is a counterpart divided by the probability that it is spurious -- must be independent of the dimensionless distance $r$ (see Eq. \ref{eq:r}). It is similar to the term  $(1-\theta)/\theta$ used in \citet{1977A&AS...28..211D}. However, $p(cp)/p(spur)$ may depend on the nature of the underlying X-ray source population (e.g. stars, AGN) and may thus vary with source properties such as magnitude, optical colour, or flux ratios. In order to obtain a $LR$ similar to that used in \citet{2007ApJS..172..353B} for instance, we would need to consider an additional parameter $q(m)$ describing the variation of $p(cp)/p(spur)$ with the magnitude (or any other relevant property) of the candidate counterpart. Alternatively, $q(m)$ may be replaced by another term such as that playing the role of $B_{phot}$ in \citet{2008ApJ...679..301B}. An $R(LR)$ histogram can be built from the $N_{cand}(LR)$ and $N_{spur}(LR)$ histograms made using the method explained in the previous paragraph. If the ratio $p(cp)/p(spur)$ were independent of source properties, $R(LR)$ could be fitted with Eq. (\ref{eq:r_ideal}) using only one free parameter $a=p(cp)/p(spur)$. Including a term $q(m)$ in $LR$ with $N_{\Delta m}$ bins of magnitude, requires us to build $N_{\Delta m}$ $R(LR,\Delta m)$ histograms and fit each of them with functions in Eq. (\ref{eq:r_ideal}) having different $a$ parameters. However, in general the lack of statistics does not allow us to do so.

The $R(LR)$ histogram can then be seen as the sum of $N_{\Delta m}$ $R(LR,\Delta m)$ histograms and consequently can be modelled by the function
\begin{equation}
    R_f(LR) = \sum\limits_{i=1}^{N_{\Delta m}}  \frac{b_i}{1+a_i\frac{1}{LR}} \mbox{ , with } b_i=\frac{N_{cand_{\Delta m_i}}}{\sum\limits_{j=1}^{N_{\Delta m}} N_{cand_{\Delta m_j}}} \ ,
\end{equation}
where $N_{cand_{\Delta m_i}}$ is the total number of entries in histogram number $i$.

In practice, the histograms are not binned according to $LR$ but to $\log_{10}(LR)$. Best fits were obtained using the function
\begin{equation}
    R_f(x=\log_{10}(LR)) = \frac{\sum\limits_{i=1}^{N_{\Delta m}} b_i' 10^{(1-i)x}   }{1+\sum\limits_{i=1}^{N_{\Delta m}} a_i' 10^{-ix}} 
\end{equation}
with $N_{\Delta m}=3$, i.e. 6 free parameters.

The fit is performed using a Levenberg-Marquard algorithm. We compute the same number of $LR$ and construct and fit the same number of $LR$ histograms as there are magnitude bands in the SDSS.

\subsection{Computing probabilities of identification in the general case \label{sec:proba_of_id}}

We now extend the Bayesian approach to X-ray sources having $N_{cand}$ candidates. We assume that at most one association is real. This assumption should be fulfilled in our case for at least two reasons. First, we only consider point-like X-ray sources. This condition decreases the probability that the detection results from two distinct unresolved sources blended in the XMM-Newton beam. Second, 95\% of the XMM-Newton sources matching a SDSS entry with a probability higher than 90\% have a 0.5-2.0\,keV flux higher than 1.65\,$\times$\,10$^{-15}$\,\ergcms. At this flux, source confusion is of the order of a few percent only \citep{cappelluti2009}. The corresponding source density of $\sim$\,500\,deg$^{-2}$ \citep{cappelluti2009} is well below the value of 2000\,deg$^{-2}$, above which simulations show that source confusion becomes important \citep{loaring2005}. Similar conclusions can be drawn for the hard (2-12\,keV) sources.

Let us consider $N_{cand}+1$ hypotheses:
\begin{itemize}
    \item $H_{cp_i}$: the i$^{th}$ optical source is the counterpart
    \item $H_{spur_{all}}$: there is no counterpart.
\end{itemize}
Then the Bayesian probability that the i$^{th}$ source is the counterpart knowing $r_j$, $j \in [1,N_{cand}]$ is
\begin{equation}
    P_{id,i} = R'_i = p(H_{cp_i}|r_1 \cap \dots \cap r_{N_{cand}})  = \frac{LR_i \frac{ p(H_{cp_i})}{p(H_{spur_{all}})} }{ 1+\sum\limits_{j=1}^{N_{cand}} LR_j\frac{p(H_{cp_j})}{p(H_{spur_{all}})}} \ .
    \label{eq:ri}
\end{equation}
If $\frac{p(H_{cp_i})}{p(H_{spur_{all}})}=\frac{p(cp_i)}{p(spur_i)}$, Eq. (\ref{eq:ri}) leads to the formula below, obtained following the \citet{2000ApJS..131..335R} prescription
 \begin{equation}
   P_{id,i} = R'_i = \frac{\frac{R_i}{1-R_i}}{1+  \sum_{i=1}^M \frac{R_i}{1-R_i}} \ ,
   \label{eq:rsun}
\end{equation}
With Eq. (\ref{eq:r_ideal}), we easily show that $R/(1-R) = LR p(cp)/p(spur)$. Computing $p(H_{cp_i})$ and $p(H_{spur_{all}})$ normalising the terms $p(cp_i)$, $p(spur_i)$ as \citet{2000ApJS..131..335R} do to construct $P_{id,i}$ from $R_i$, we obtain the equality $\frac{p(H_{cp_i})}{p(H_{spur_{all}})}=\frac{p(cp_i)}{p(spur_i)}$. We thus apply Eq. (\ref{eq:rsun}) to compute the final probabilities of identification.

Each candidate possesses as many reliabilities as there are magnitude bands in the SDSS. The $R_i$ we consider in the final probability of identification formula are for each source the best of all photometric bands.

\section{Observation grouping}\label{obsgrouping}

XMM-Newton EPIC sources are correlated FOV by FOV, ie, observation by observation. In order to tail off count-rate noise on FOV $LR$ histogram bins without sacrificing resolution, we have to increase count statistics. We therefore stacked data from similar FOV:
\begin{itemize}
    \item We split into two groups XMM-Newton FOV with different systematic errors on position: 0.35'' or 1.0''.
    \item Observations of the LMC and SMC regions are set apart.
    \item Because they presumably share objects of same nature and same patterns of logN-logS relation, observations are
	  grouped according to their galactic latitude.
\end{itemize}

As mentioned above, the relation between reliability and likelihood ratio depends on the overall properties of the X-ray populations present in the optical sample. In addition to galactic latitude, we also tested whether the X-ray flux could significantly modify the shape of the $R(LR)$ curves. Splitting further 2XMMi sources into groups of medium (10$^{-14}$--10$^{-13}$\,\ergcms) and faint (10$^{-15}$--10$^{-14}$\,\ergcms) 0.2-12 keV flux ranges does not change the probabilities of identification by more than $\sim$ 3\% in most cases. The only noticeable difference is for faint sources with identification probabilities below 50\%, which tend to show even lower identification probabilities by as much as 15\%. We felt, however, that since the effect is relatively modest and only affects sources for which the significance of the identification is rather low, priority should be given to the gathering of sufficiently large subsamples. We thus did not consider any X-ray flux dependency in the final implementation.

\section{Results of the 2XMMi-SDSS DR7 cross-correlation}

\begin{figure*}
    \begin{tabular}{ccc}
    \includegraphics[width=0.3\textwidth]{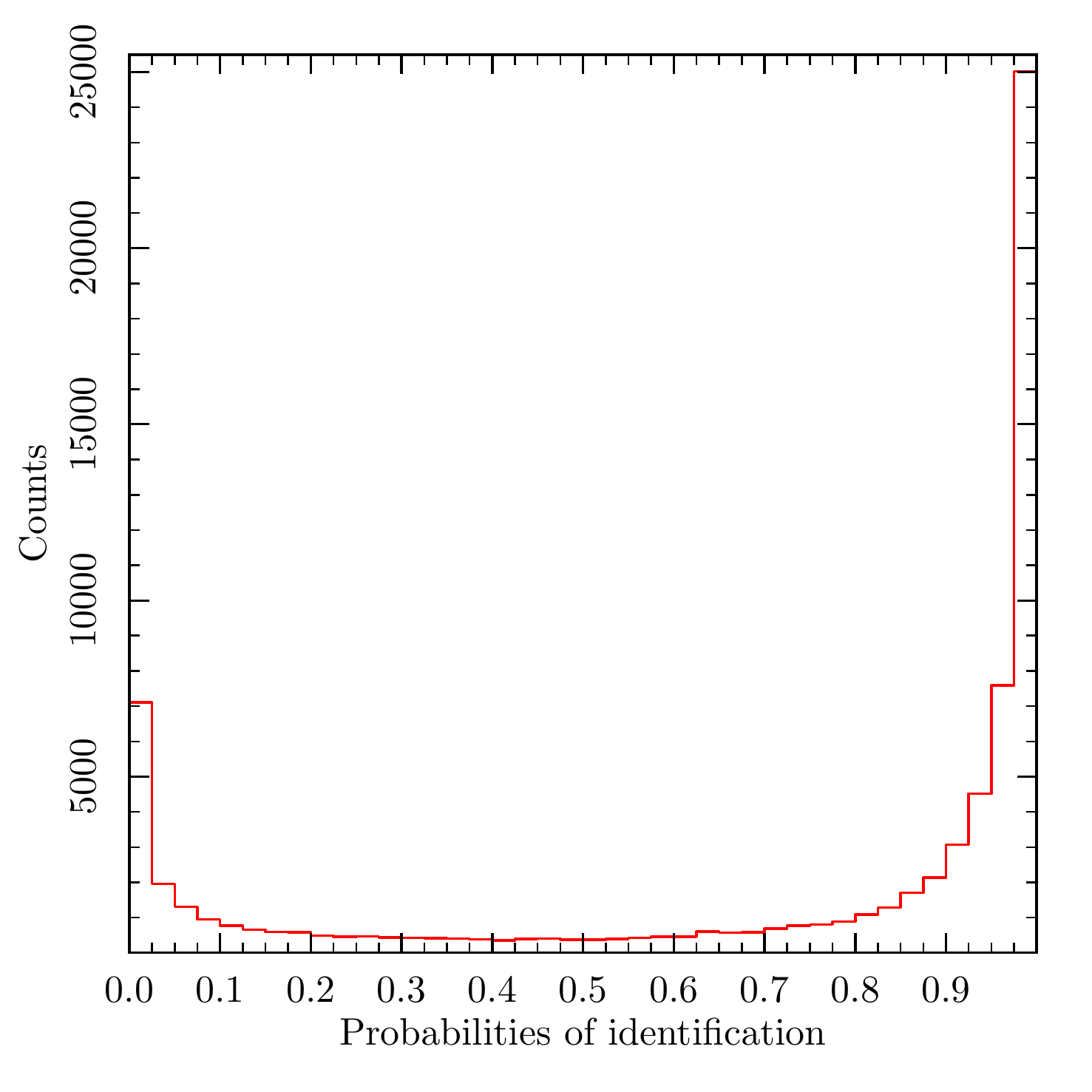} &
    \includegraphics[width=0.3\textwidth]{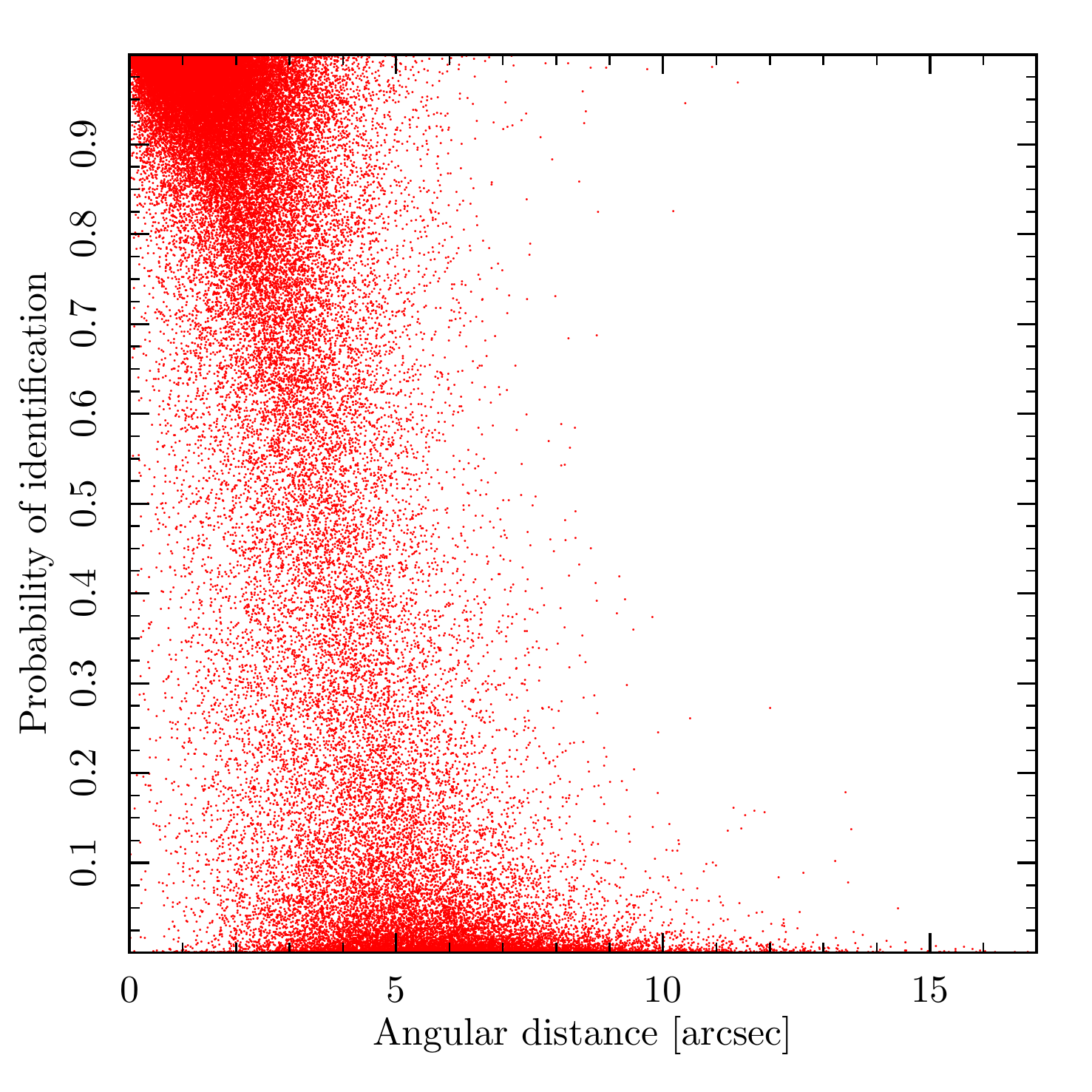} &
    \includegraphics[width=0.3\textwidth]{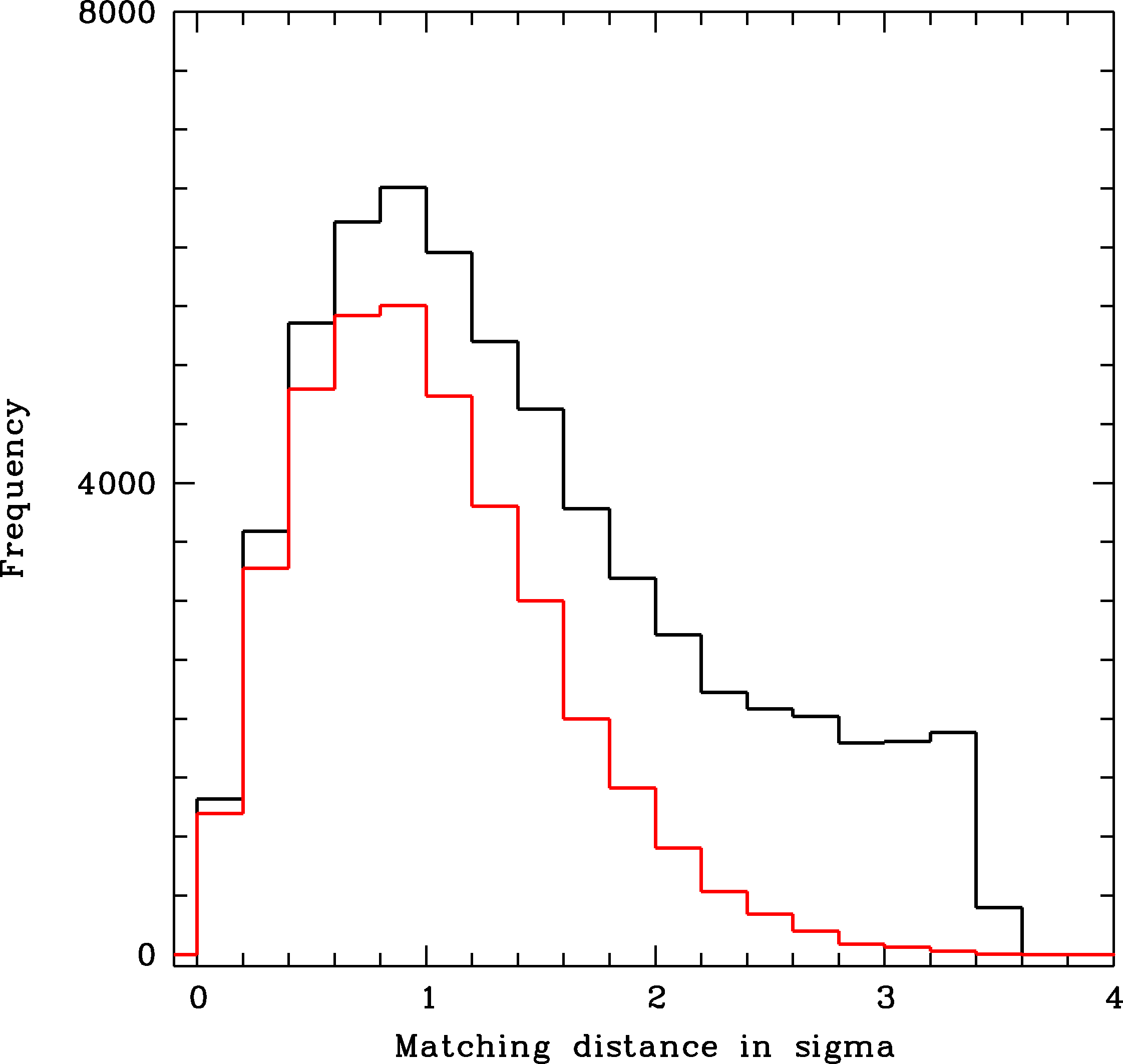}\\
    \end{tabular}
    \caption{Left: Histogram of the individual probabilities of identification. Centre: probabilities of identification versus matching distances. Right: distribution of the distance of SDSS candidates to 2XMMi sources expressed in units of the combined 2XMMi + SDSS positional error; black = all matches, red = identification probabilities $\geq$ 90\%.}
    \label{fig:proba_id}
\end{figure*}

A total of 1337 XMM-Newton FOV hold at least one source with a SDSS counterpart candidate within the combined 3\,$\sigma$ search radius. These 1337 FOV contain 95\,452 detections, corresponding to 73\,636 unique 2XMMi sources. The cross-correlation of the 2XMMi catalogue with the SDSS DR7 leads to 72\,169 ``associations'' involving 45\,727 and 55\,726 unique 2XMMi and SDSS DR7 sources respectively. This first number represents 20\% and 62\% of the unique sources available in the entire 2XMMi catalogue and in the 1337 FOV respectively. The distribution of the number of SDSS DR7 candidates by unique 2XMMi sources is given in Table \ref{tab:nbr_cand_dist}, and the main properties of the distribution of the probabilities of identification and of their cumulative values are given in Table \ref{tab:res_stat}. We define the sample completeness as the fraction of 2XMMi sources having an individual probability of identification in SDSS above a given cutoff relative to the total number of 2XMMi sources with SDSS counterparts. In a similar manner, sample reliability is the fraction of non-spurious associations among 2XMMi sources having an individual probability of SDSS association above a given threshold.  A total of 7\,740 unique 2XMMi sources have several SDSS DR7 candidates. In this sub-sample, there are 896 and 2\,672 2XMMi sources for which the candidate with the highest identification probability is not the nearest and the brightest SDSS DR7 candidate respectively. 

The left panel of Fig. \ref{fig:proba_id} shows the distribution of the individual SDSS source identification probabilities. Most SDSS entries found within the combined 3$\sigma$ search radius from the 2XMMi source have a high likelihood to be the true optical counterpart. The small tail of very low identification probability objects reflects the expected rising contribution of SDSS entries unrelated to the X-ray source at large matching distances. Most SDSS entries with identification probability higher than $\sim$ 90\% are found less than 3 arcsec from the X-ray position (Fig. \ref{fig:proba_id}, centre and right panel). The rather wide spread of the 2XMMi positional errors accounts for the scatter affecting the distances at which high-probability SDSS sources are found from the X-ray position.

Expressed in terms of combined 2XMMi + SDSS errors, the distance distribution shown in the right panel of Fig. \ref{fig:proba_id} \ follows the usual shape of a Rayleigh distribution. Fitting this histogram with a Rayleigh function plus a linear component, we obtain $\sigma_r=0.865$ for the Rayleigh curve parameter, 0.178 for the linear slope and $R=0.675$ for the ratio between the total number of real associations and the total number of spurious ones within the search radius. Fitting separately distance histograms of the sources whose positions were corrected by {\em eposcorr} and uncorrected ones leads to $\sigma_r=0.856$, $R=0.698$ and $\sigma_r=1.013$, $R=0.50$ respectively. All errors on $\sigma_r$, on the slope and on the $R$ values are about 0.003. Keeping only the best candidate for each unique XMM source, we obtain R=0.81, which is consistent with the value of 79.4\% given in Table \ref{tab:res_stat}. The origin of this small apparent overestimate ($\sim$ 14\%) of the positional errors of {\em eposcorr} corrected sources is so far unclear. In any case, the effect of this slight change on the identification probabilities is small. The global effect is to slightly decrease the probabilities of SDSS entries matching at large distances and to somewhat increase the probabilities of those located close to the X-ray source. 

\begin{table}
    \centering
    \caption{ Number of unique 2XMMi sources (Nx) with Nc SDSS DR7 candidates.
    }
    \begin{tabular}{l|c|c|c|c|c|c|c|c|c|c}
	Nc &       1 &    2 &    3 &   4 &  5 & 6  & 7  & 8 & $>$8 \\
	\hline
	Nx  & 37\,988 & 6059 & 1196 & 317 & 96 & 47 & 12 & 4 & 8\\
    \end{tabular}
    \label{tab:nbr_cand_dist}	   
\end{table}

\begin{table}
    \caption{Cross-correlation statistics.}
    \begin{tabular}{c|r|r|r|r|r|r}
	\multicolumn{1}{c|}{ } & \multicolumn{6}{c}{ Probability of identification cutoffs}\\
	\cline{2-7}
	\multicolumn{1}{c|}{id} & 0.0 & 0.5 & 0.7 & 0.8 & 0.9 & 0.95 \\
	\hline
	\# det\_id & 60\,567 & 53\,347 & 49\,527  & 46\,387 & 40\,193 & 32\,610 \\
	\# src\_id & 45\,727 & 39\,839 & 36\,943  & 34\,605 & 30\,055 & 24\,327 \\
        R          & 79.4\%  & 91.7\%  & 94.7\%  & 96.2\%  & 98.0\%  & 99.0\%  \\
        C          & 100.0\% & 96.8\%  & 92.1\%  & 87.5\%  & 77.2\%  & 63.2\%   \\
        Frac X     & 62.1\% & 54.1\%  & 50.2\%  & 47.0\%  & 40.8\%  & 33.0\%   \\
    \end{tabular}
    \note{Number of associations with all XMM detection (\emph{det\_id}) and all unique sources (\emph{src\_id}) with a probability of identification above a given probability of identification. If an X-ray source has several candidates, we only keep the one with the best probability of identification. Sample reliability (R) and sample completeness (C): we only consider the best match for each unique XMM sources having at least one counterpart. Frac X is the expected fraction of X-ray sources with a counterpart in the SDSS DR7 catalogue.}
    \label{tab:res_stat}      
\end{table}

The practical implementation is described in Appendix \ref{implementation}. Whenever optical data are used, we discard SDSS entries with recorded magnitudes fainter than 22.2 in any of the photometric bands considered. Indeed, objects with magnitudes higher than 22.2 tend to have smaller photometric errors than brighter ones, clearly indicating that SDSS photometric uncertainties and perhaps also mean values are not reliable at faint flux. We also ignored all SDSS entries having one of the following flag set: BLENDED, DEBLENDED\_AS\_MOVING, SATURATED, INTERP\_CENTER, EDGE, SATUR\_CENTER, PSF\_FLUX\_INTERP in order to ensure the best photometric quality. Unless specified otherwise, we will hereinafter only consider optical identifications with a probability larger than 90\%. This threshold applies to both the spectroscopically identified sample and to the general photometric sample and corresponds to an overall sample purity of 98\% (see Table \ref{tab:res_stat}).

\section{Building an identified sample}\label{buildingls}

One of the important task given to the SSC is the statistical identification and classification of all X-ray sources discovered in the wide field of view of the EPIC cameras. The statistical determination of the nature of any given 2XMMi source will first rely on the assessment of the reliability of its association with candidate counterparts at other wavelength. The description of this important step and of its results are the goals of the present paper. 

On the other hand, the subsequent classification stage requires the knowledge of the parameter space occupied by the various groups of astrophysical sources using a ``learning sample''. 
Therefore, the cross-correlation method presented here allows us to select in a clean and statistically controlled manner the best optical counterparts to 2XMMi sources and constitutes the first mandatory step towards building a reliable learning sample, which can be later used to define source classification schemes using advanced statistical methods. Eventually, the classification method, either supervised or not, will provide the most likely nature of the 2XMMi source (e.g. star, AGN, etc..) with for some methods, an estimate of the probability of the classification. First attempts to classify 2XMMi sources in two classes (stars and  extragalactic) have been presented in \cite{pineau2009} and are now implemented in the XCat-DB for the DR3 of the 2XMM catalogue.  

Yet it is also well known that a reliable classification can only be achieved when the corresponding learning sample covers the parameter space spanned by the group of objects to identify as evenly as possible, see e.g. \citet[][]{white2008} or \citet{richards2004}. Being aware of this important requirement, the SSC has designed a general optical identification programme able to explore the widely diverse natures of the X-ray emitting objects discovered in the XMM source catalogues.  Several wide field identification campaigns are currently conducted at various X-ray flux levels and galactic latitudes, which all aim at building completely identified source samples. The nature of the high $b$ population is the scope of four distinct projects. The bright part is studied by the Bright Sources Survey \citep[XBS or BSS,][]{dellaceca2004,caccianiga2008}. The XMM-Newton Medium Sensitivity Survey \citep[XMS,][]{barcons2002,carrera2007,barcons2007} and the XMM-2dF Wide Angle Survey \citep[XWAS,][]{tedds2006} investigate the properties of medium flux sources. The faintest source population is the scope of the Subaru/XMM-Newton Deep Survey \citep[SXDS,][]{useda2008}. Finally, the Galactic plane area is covered by the XMM-SSC Galactic Plane Survey \citep{motch2005,motch2010}.

The first step towards building a sample of 2XMMi sources of known astrophysical nature was to select X-ray sources with reliable SDSS DR7 spectroscopic counterparts of a known class (i.e., with the {\it specClass} attribute pointing to an astrophysical object). 
For our purpose, the three most important groups of spectroscopic SDSS targets are the sample of quasar candidates defined by \cite{richards2004}, the main galaxy sample described in \cite{strauss2002} and all stars belonging to the legacy survey and to the Sloan Extension for Galactic Understanding and Exploration programme \citep[SEGUE, ][]{yanny2009}. The AGN sample is mostly a classically UV-excess (UVX) selected sample to which is added a small number of redder targets appearing as likely high redshift QSOs. The galaxy sample is less biased because it is only selected on brightness related criteria in the $r$ band. Stars from the legacy survey were mostly selected on the basis of their extreme colours. Among them, red dwarfs and CVs are the most likely to match 2XMMi sources. The SEGUE programme opens new areas at lower galactic latitudes, and its spectroscopic target selection aims at covering all spectral types. 

We therefore extracted the SDSS spectroscopic catalogue accessible via CasJob, and following the SDSS spectral class scheme, define the classes: stars, galaxies, AGN and X-ray accreting binaries. We list below the origin of the different groups of identified sources:

 \begin{itemize}
  \item Stars~: i) 2XMMi/SDSS associations having the {\it specClass} attribute set to 1 or 6 and ii) the sample of stars coming from the kernel density classification (see Sect. \ref{xstars}).
  
  \item Accreting binaries~: i) objects in the Downes catalogue of cataclysmic variables \citep{2001PASP..113..764D}. We used here the 2006 version, which contains many SDSS discoveries; ii) the Ritter catalogue of cataclysmic variables \citep{2003A&A...404..301R} and iii) the Ritter catalogue of LMXRBs \citep{2003A&A...404..301R}.
  
  \item Galaxies~: 2XMMi/SDSS associations with a probability of identification $>$ 0.80 and with the {\it specClass} attribute set to 2.
  
  \item AGN: i) sources from the V\'eron catalogue \citep{2006A&A...455..773V} and ii) SDSS DR7 objects associated with a 2XMMi source with a probability of identification $>$ 0.80 and having the {\it specClass} attribute set to 3 or 4 (QSO or high $z$ QSOs). 
  
\end{itemize}

We use the range of X-ray luminosity to define several groups of active galaxies and consider all extragalactic objects as a single class. In particular, we do not make any formal distinction between QSO and AGN\footnote{A large fraction of these AGNS are found the V\'eron catalogue \citep{2006A&A...455..773V}, which is based on the DR4. Among the 1290 SDSS DR7 spectroscopic QSOs entries, 836 are also present in the V\'eron catalogue and are classified QSOs or AGN according to their absolute blue magnitude M{\sc B}.}. An X-ray source associated with both a star and an accreting binary was flagged as an accreting binary. We applied the same rules for star--AGN and binary--AGN pairs of apparently conflicting nature. 

We added sources identified in the XBS and XMS SSC surveys and with SDSS counterparts. For the XMS, we considered their sources with classes NELG, BLAGN, and BLLac as AGN. For the XBS, AGN2, AGN1, BLLac and elusive AGN were assigned the general AGN type. 

QSO2s candidates taken from \cite{zakamska2003} and \cite{reyes2008} as well as a handful of X-ray selected objects (see Sect. \ref{qso2} below) having a reliable match in the 2XMMi catalogue were added to the identified sample. 

\subsection{The stellar identified sample}\label{xstars}

Building a clean stellar sample turned out to be more difficult because most stellar sources detected in X-rays have optical SDSS magnitudes brighter than 15 mag and are flagged as saturated. Furthermore, the SDSS DR7 spectroscopic database provided only few cross-matches with acceptable properties (i.e. non-saturated and probabilities of identification higher than 90\%). Therefore, in order to enlarge the stellar sample, we applied a classification method allowing us to identify stars on the basis of their multi-colour properties. 

We performed a kernel density classification \citep[KDC,][]{richards2004} on all spatially unresolved (cl=6) SDSS candidates. 
This selection returns 10\,533 SDSS sources with a correlation in the 2XMMi catalogue.  The classification only uses the four colours $u-g$, $g-r$, $r-i$ and $i-z$ as parameters. The learning sample used for this classification consists of two classes: star and QSO, since we only consider point-like objects in the optical. It has been built from all unresolved SDSS sources, independently of their association with a 2XMMi entry. We only retained good quality detections (i.e. no flag SATURATED, BLENDED, DEBLENDED\_AS\_MOVING, INTERP\_CENTER, EDGE, SATUR\_CENTER or PSF\_FLUX\_INTERP set) that were spectroscopically identified in the DR7. The data have been retrieved from the DR7 database with CasJob. The stellar sample contains 67\,269 sources flagged by the SDSS {\it specClass} attribute as star (STAR or STAR\_LATE) and therefore also contains CVs and WDs. The non--stellar sample has 75\,248 sources flagged by the SDSS {\it specClass} attribute as QSO (QSO) or high-redshift QSO (HIZ\_QSO) plus 253 sources flagged by {\it specClass} as galaxy (GALAXY). For simplicity we call the non--star sample AGN sample below.

Estimates of the probability densities were computed using a fixed bandwidth kernel smoothing. The kernel applied uses the Epanechnikov profile and the bandwidth was chosen to be equal to 0.2 mag.
Table \ref{tab:classif_stats} lists the results of the self-check of the learning sample, i.e., the results of the classification method applied to the learning sample only. 

\begin{table}
 \centering
 \caption{Results of the classification method applied to the learning sample.}
 \begin{tabular}{l|rr}
   org\verb+\+assign &    Star  &   AGN     \\ \hline
   Star              &  96.89\% &  3.11\%   \\
   AGN               &   1.62\% & 98.38\%   \\
 \end{tabular}
 \label{tab:classif_stats}
\end{table}

The prior probability p(star) has been set to 0.25 and so p(AGN) to 0.75 as a result of iterative kernel density classifications converging to this relative number of stars and AGN in the SDSS/2XMMi learning sample. In order to select SDSS/2XMMi identifications with the best chance to be normal stars, we removed 13\% of all SDSS entries classified as stars, but falling in low-density regions of the parameter space (i.e. far from the centre of the stellar multi-colour locus) and thus prone to be doubtful cases such as binaries, unidentified cataclysmic variables, or even mis-identified AGN. 

The star/AGN classification has been made according to the optical properties of the spectroscopically identified SDSS objects. However, by construction, the density distribution in colours of the SDSS/2XMMi sample is not likely to follow that of the non X-ray emitting SDSS objects and can thus lead to some biases. For instance, there may be a considerable overdensity of non-X-ray emitting stars in some part of the 4-d colour diagram where most objects classified as AGN appear to be strong X-ray sources. This problem indeed occurs in the region covered by the AGN branch, where there is some overlap with A stars. Although some A stars do emit X-rays for debated reasons, not all do. We thus removed from the SDSS/2XMMi learning sample all classified stars with a $u-g$ colour of less than 1.2 \citep[values taken from][]{covey2007}. 

The final stellar X-ray sample arising from the KDC contains 636 unique entries with a classification probability higher than 99.7\% (3 Gaussian $\sigma$). However, only 549 of these matches have a probability of identification with a 2XMMi source higher than 90\% and were therefore entered in the final identified sample.

We also checked that the stellar SDSS/2XMMi sample adhered to the stellar locus derived by \citep{covey2007} using synthetic photometry in the 4-D colour space. The agreement is good, apart for the reddest stars of spectral type later than $\sim$ M5.

\subsection{The final identified sample}\label{finalidsamp}

\begin{table}
\tabcolsep=3pt
    \centering
    \caption{Distribution of XMS and XBS sources in the identified samples.}
    \begin{tabular}{l|c|c|c|c|c|c}
       \multicolumn{1}{c|}{Sample} &  \multicolumn{2}{c|}{Galaxy} & \multicolumn{2}{c}{AGN} & \multicolumn{2}{c}{QSO} \\
       \cline{2-7}
       \multicolumn{1}{c|}{ } & XMS & XBS & XMS & XBS & XMS & XBS \\ \hline
       All                    &  1  &  0  &  7  &  21 &  49 &  95 \\ 
       Final                  &  1  &  0  &  5  &  14 &  37 &  85
    \end{tabular}
    \note{the ``Final'' sample only contains associations with a probability of identification $>$ 0.9, with SDSS sources unsaturated, not blended, ..., and where all magnitudes $<$ 22.2 mag.} 
    \label{xidsamp}
\end{table}

\begin{table}
    \caption{Origins of the objects of each type in the final identified sample.}
    \begin{minipage}{8.3cm}
    \centering
    \begin{tabular}{l|c|c|c|c|c|c|c}
       \multicolumn{1}{c|}{Sample} & \multicolumn{2}{c|}{Star} & \multicolumn{3}{c|}{Accreting Binary} & \multicolumn{2}{c}{Extragalactic} \\
       \cline{2-8}
       \multicolumn{1}{c|}{ } &  DR7 & KDC & Dow.\footnote{Downes catalogue of CV.} & R1 \footnote{Ritter catalogue of CV.}& R2\footnote{Ritter catalogue of LMXB.} & DR7 & Veron \\ \hline
       Final\footnote{Note that individual sources can be listed in several catalogues.}                  &  8   & 541 & 29   & 22 & 2  & 2021& 1524  \\ 
    \end{tabular}\par
   \vspace{0.2\skip\footins}
   \renewcommand{\footnoterule}{}
   \end{minipage}
    \label{idsamp}
\end{table}

\begin{table}
    \centering
    \caption{Distribution of number of unique 2XMMi entries with types in the final identified sample.}
    \begin{tabular}{l|c|c|c}
       Sample & Star & Accr. Binary & Extragalactic \\ \hline
       Sub    & 549  &  26    & 2336 \\
    \end{tabular}
    \label{finallearningsample}
\end{table}

The origin and distribution of the various classes of identified objects in the final sample are listed in  Tables  \ref{xidsamp}, \ref{idsamp}, and  \ref{finallearningsample}. Most of the extragalactic identified sample comes from the DR7 spectroscopic catalogue through the V\'eron catalogue, while the vast majority of X-ray active stars are actually extracted from the KDC source classification. A large fraction of the identified accreting binaries (mostly cataclysmic variables) also come from follow-up SDSS discoveries through the Downes catalogue.

Finally, for all AGNs with a spectroscopic redshift we compute the observed X-ray luminosity using
\begin{equation}
    L_{abs} = 4\pi F \frac{(1+z)^2}{(1+z)^{2-\gamma}} \left( \frac{c}{H_0} \int_0^Z \frac{1}{\sqrt{\Omega_m(1+z)^3+\Omega_\Lambda}} dz \right)^2 \ ,
    \label{eq:l_abs}
\end{equation}
where $F$ is the 0.2-12 keV X-ray flux in erg\,s$^{-1}$cm$^{-2}$, $H_0=73$\,km\,s$^{-1}$\,Mpc$^{-1}$, $\Omega_m=0.3$ and $\Omega_\Lambda=0.7$. The photon index $\gamma$ was taken to be $1.9$.

\section{Grouping X-ray sources in parameter space}

\subsection{Sample properties and shortcomings}

\begin{figure*}
    \begin{tabular}{ccc}
    \includegraphics[angle=-90,width=0.31\textwidth,viewport=0 120 550 630]{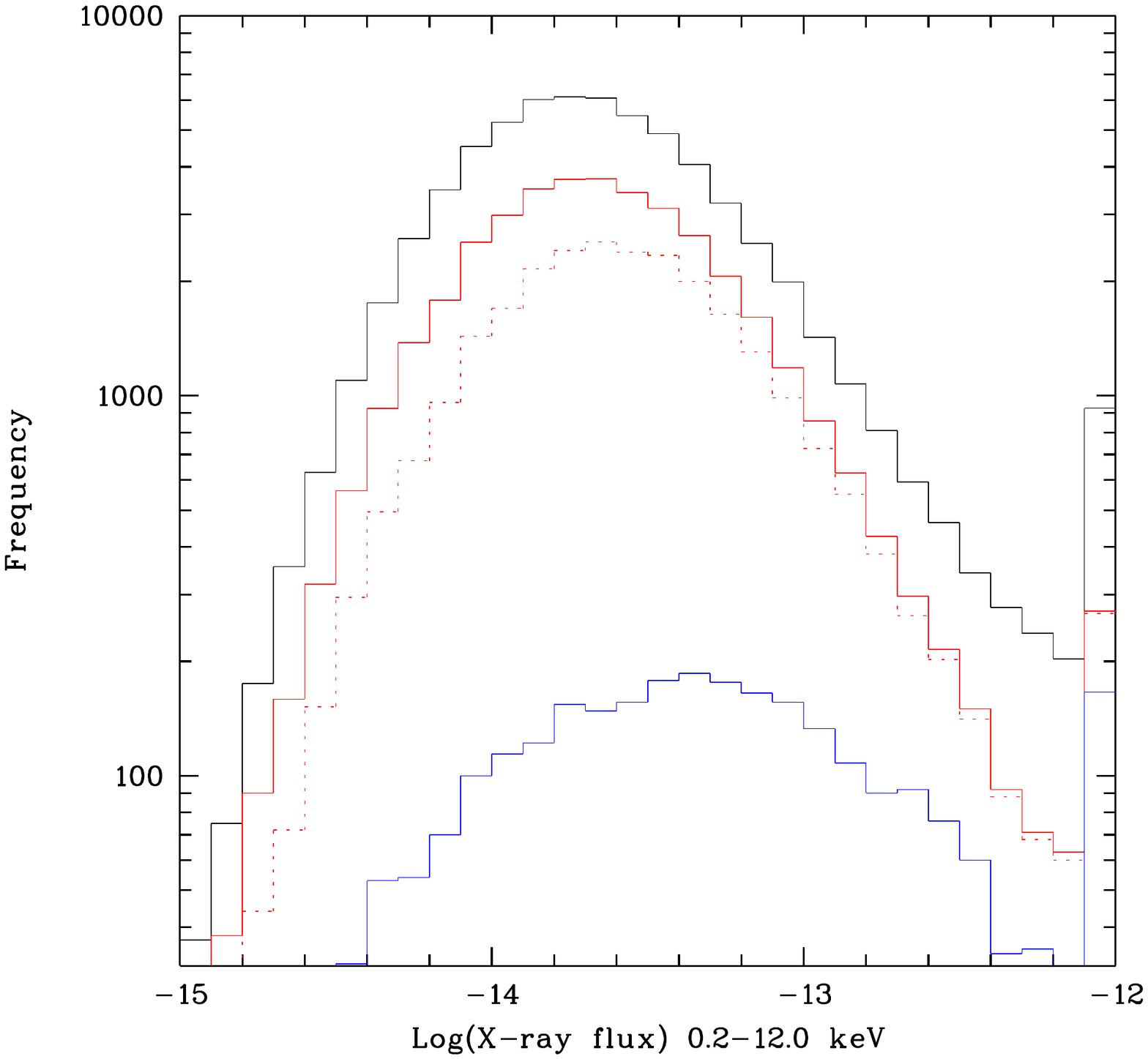} &
    \includegraphics[angle=-90,width=0.31\textwidth,viewport=0 120 550 630]{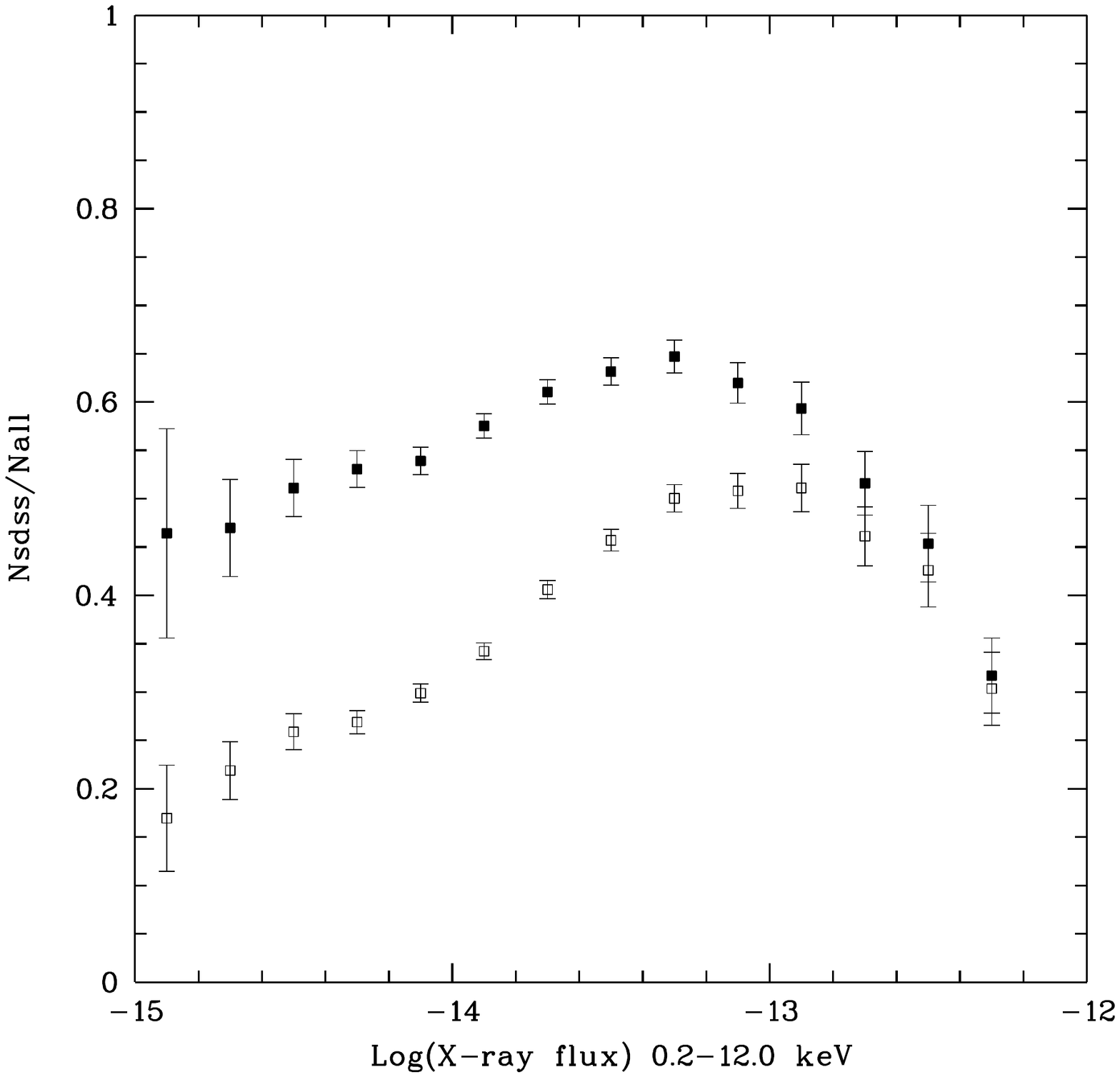} &
    \includegraphics[angle=-90,width=0.31\textwidth,viewport=0 120 550 630]{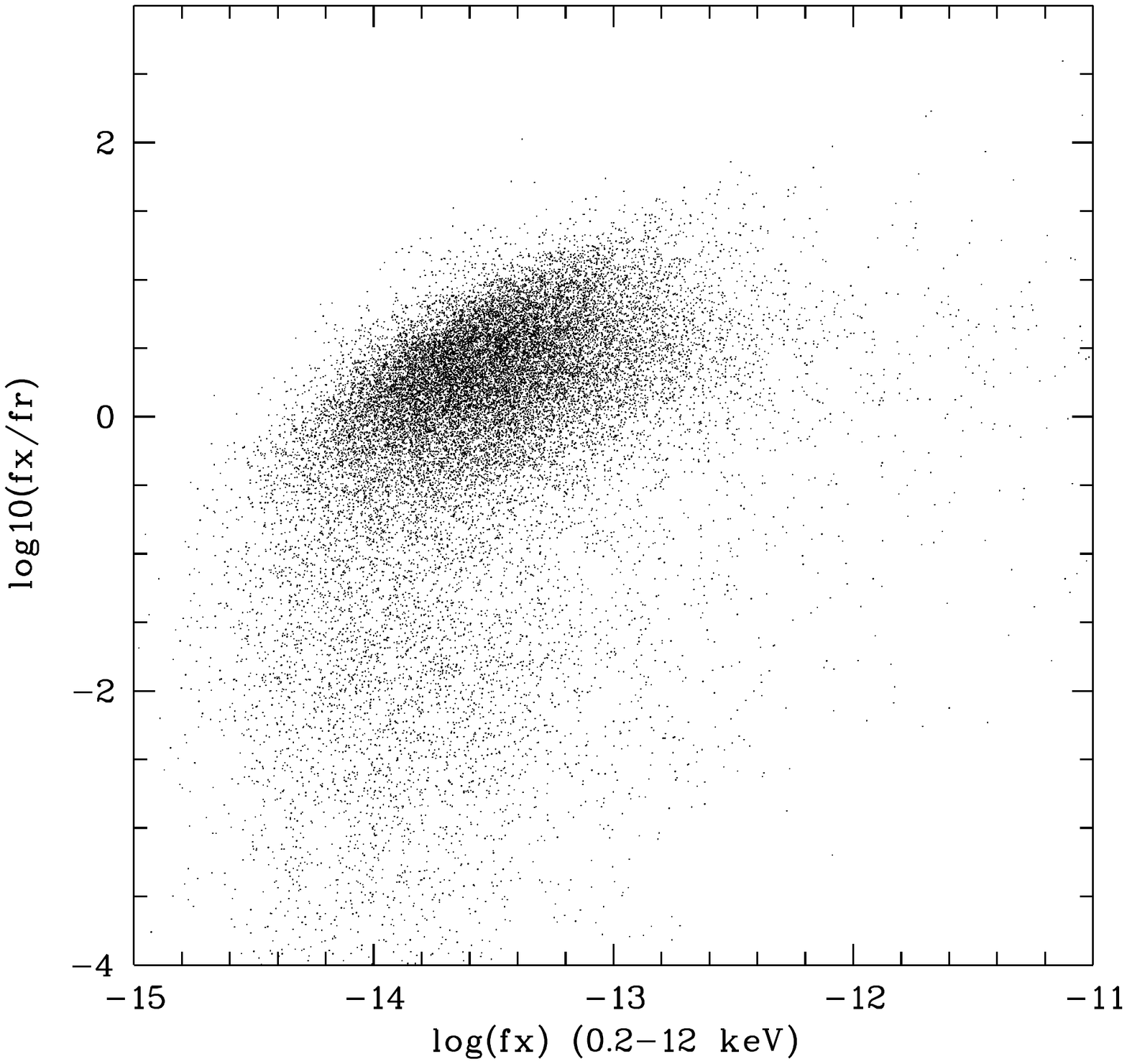} \\
    \end{tabular}
    \caption{Left panel: X-ray flux distribution. From top to bottom, black line: all X-ray sources found in 2XMMi fields overlapping SDSS DR7, red line: all X-ray sources matching a SDSS entry, red dashed line: 2XMMi/SDSS R7 correlations with individual identification probabilities $>$ 90\%, blue line: ``final'' identified sample. Centre panel: ratio of the SDSS identified to total number of 2XMMi sources. Upper curve, filled squares; all matches, lower curve, empty squares, matches with identification probabilities $\geq$ 90\%. Right panel: Variation of log(fx/fr) with X-ray flux for 2XMMi/SDSS DR7 sources with identification probabilities $\geq$ 90\% and $r$ magnitude brighter than 22.2. In all cases, the combined unique EPIC source detection likelihood is $\geq$ 6. }
\label{plot_flux_distributions}    
\end{figure*}

The left and centre panels of Fig. \ref{plot_flux_distributions} display the X-ray flux distributions of 2XMMi sources with individual identification probabilities with SDSS DR7 entries $\geq 0\%$ and $\geq 90\%$, compared to that of all 2XMMi sources present in the SDSS DR7 footprint. Many faint X-ray sources have likely counterparts in the SDSS. Figure \ref{plot_flux_distributions} also shows that the fraction of 2XMMi sources with likely SDSS identifications does not vary strongly with X-ray flux. It steadily decreases by a factor of 2 from \Fx\,$\sim$ $1\,\times\,\sim\, 10^{-13}$\,\ergscm\ to $1\,\times\,\sim\,10^{-15}$\,\ergscm . The drop in the identification rate at a flux above $\sim$ 10$^{-13}$\,\ergscm\ is probably caused by the increasing number of bright optical counterparts, which are likely to be flagged as saturated in SDSS and therefore absent from our sample. On the other hand, the shape of the decline of the SDSS identified fraction with decreasing X-ray flux (centre panel) as well as the observed distribution of the $f_x/f_{opt}$ ratio with X-ray flux (right panel) are both consistent with populations of X-ray sources with a weakly varying distribution of $f_x/f_{opt}$ ratios with X-ray flux. In other words, comparing the SDSS identified sample with the total sample does not reveal evidence of strong evolution of $f_x/f_{opt}$ with redshift. A comparable conclusion was reached in Sect. \ref{obsgrouping} based on the weak dependency with X-ray flux of the reliability / likelihood ratio relations. It may thus be possible to extrapolate the properties of the 2XMMi/SDSS DR7 photometric identified sources to somewhat fainter X-ray and optical fluxes. 

The situation of the ``final'' spectroscopic identified sample clearly differs. Its X-ray flux distribution strongly differs from that of the 2XMMi/SDSS DR7 photometric sample and from that of the overall 2XMMi sample. Obviously, this discrepancy arises from the higher optical brightness needed by spectroscopic observations (see left panel of Fig.\ref{plot_flux_distributions}). In addition, the choice of the spectroscopic targets results from various heterogeneous optical selection criteria and is therefore unlikely to cover all X-ray emitting objects above equally any given X-ray flux threshold. Some examples are outlined in the sections below. We therefore stress that as its stands, this identified sample cannot in any manner be used as a learning sample suitable for a statistically reliable classification of 2XMMi sources with SDSS identifications.   

However, the high number of spectroscopic SDSS matches supplemented by other identifications derived from archival catalogues allows us to build an unprecedentedly large sample of X-ray sources of known nature, enriched with accurate multicolour photometry and detailed spectral line measurements. This large collection of best quality data offers a unique opportunity to study to some extent the parameter locii occupied by the different classes of X-ray emitters. However, it also allows addressing two important issues. First, finding the most efficient physical parameters for separating different groups of X-ray sources. Second, it allows us to highlight the parameter regions not well covered by the SDSS observing strategy and therefore in need of extended spectroscopic studies.

\begin{figure*}
    \begin{tabular}{cc}
      \includegraphics[angle=-90,width=0.5\textwidth,viewport=0 50 550 650]{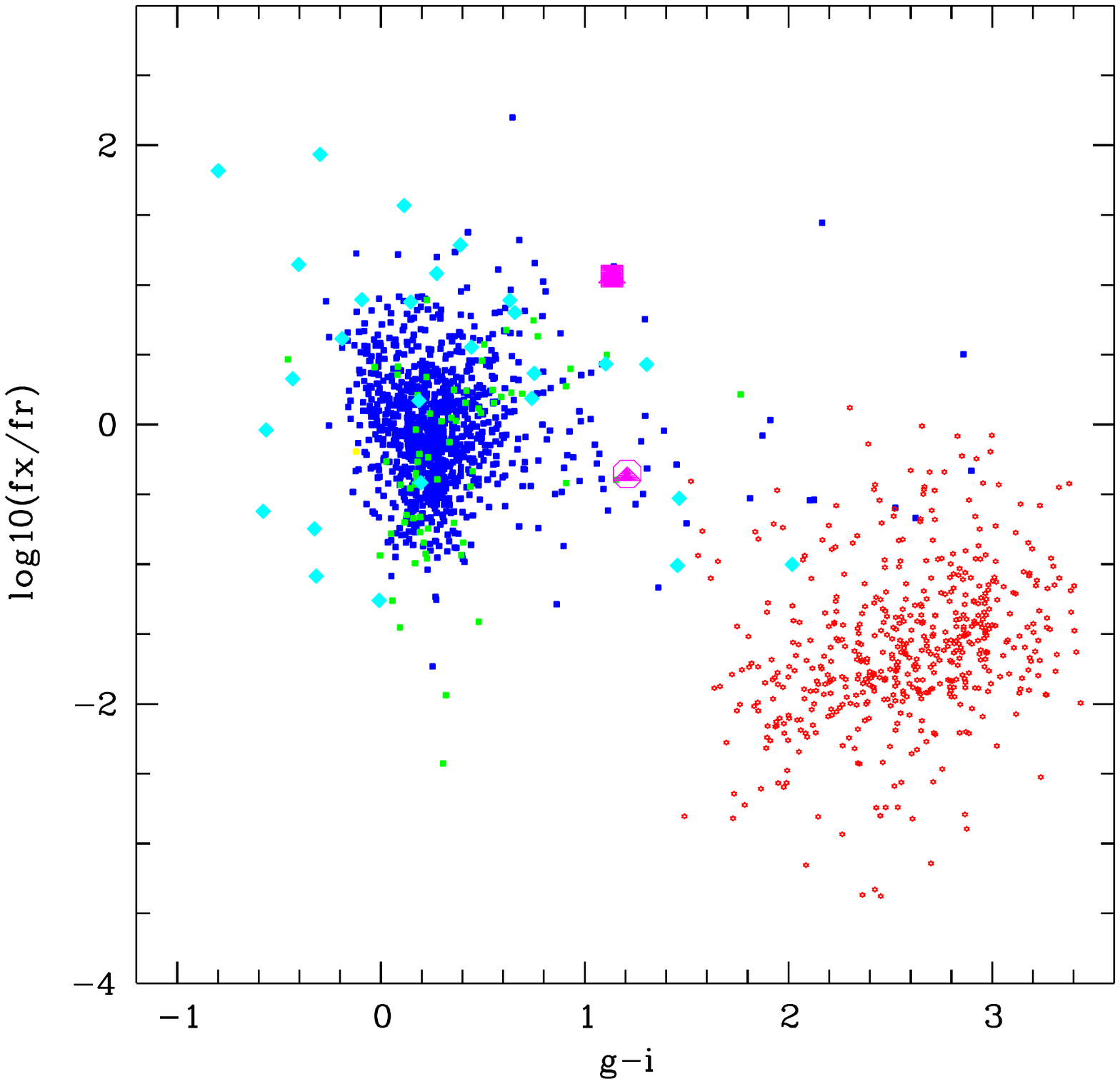} &
      \includegraphics[angle=-90,width=0.5\textwidth,viewport=0 50 550 650]{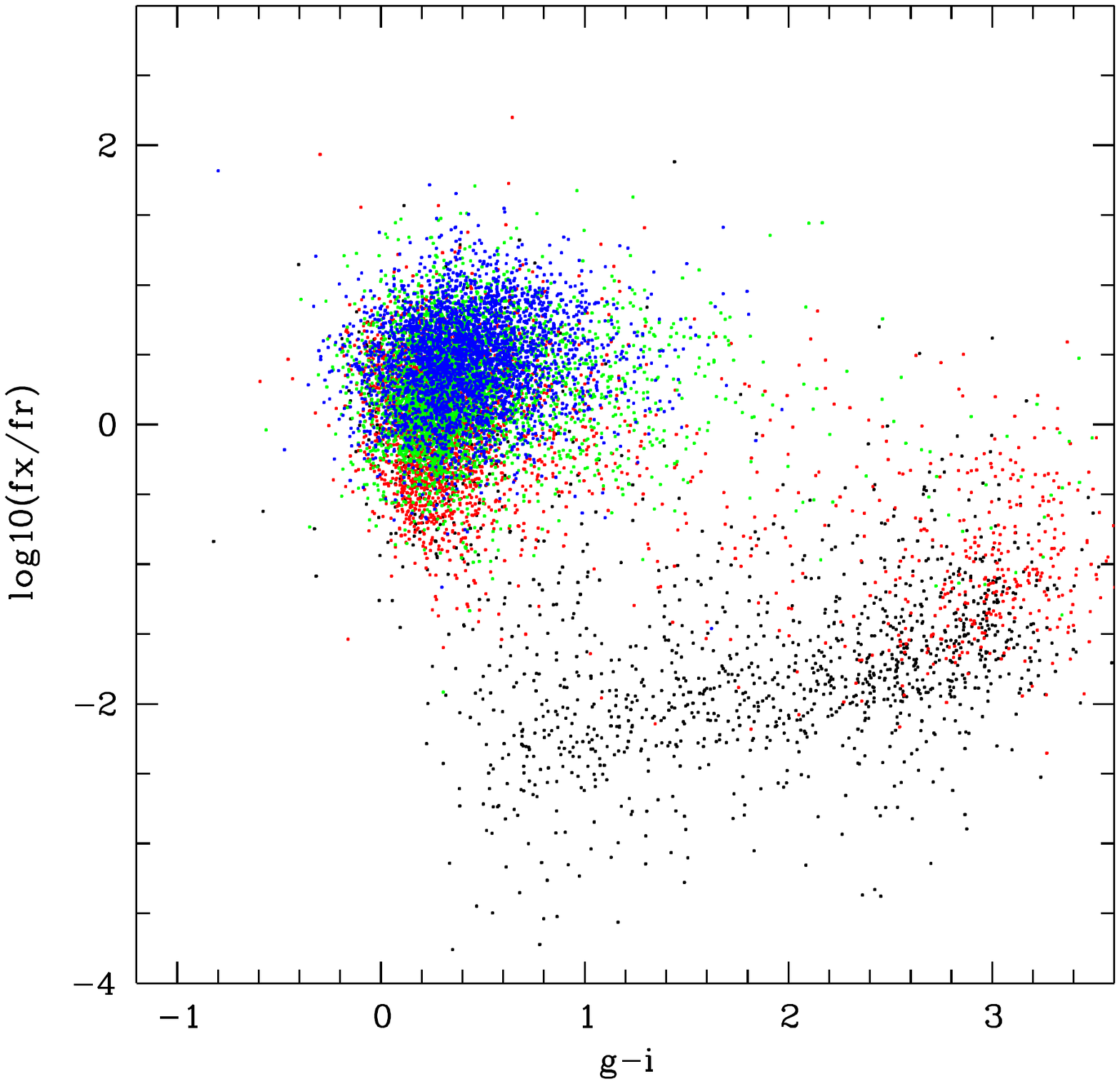} \\
    \end{tabular}        
    \caption{Distribution of spatially unresolved objects in the optical band in the $log(f_x/f_r)$ versus $g-i$ diagram. Left: the identified sample. Blue: AGNs with $log($\Lx$) \geq$ 44, green: $44 \geq log($\Lx$) \geq 42$, magenta: QSO2s
- filled squares = X-ray selected - filled triangles = optically selected - encircled = Compton Thick (see Sect. \ref{qso2}), red: stars, cyan: accreting binaries. Right: the entire SDSS photometric sample. In this case, the colour codes the range of $r$ magnitude. black: $<$18, red:18-20, green, 20-21, blue:$>$21.  We only show SDSS entries with a probability of identification with an X-ray source higher than 90\%, $g$ and $i$ magnitudes brighter than 22.2 and errors on $g-i$ $<$ 0.2.}
\label{fig:fxfo_vs_gi_pl_type}    
\end{figure*}

\begin{figure*}
    \begin{tabular}{cc}
      \includegraphics[angle=-90,width=0.5\textwidth,viewport=0 50 550 650]{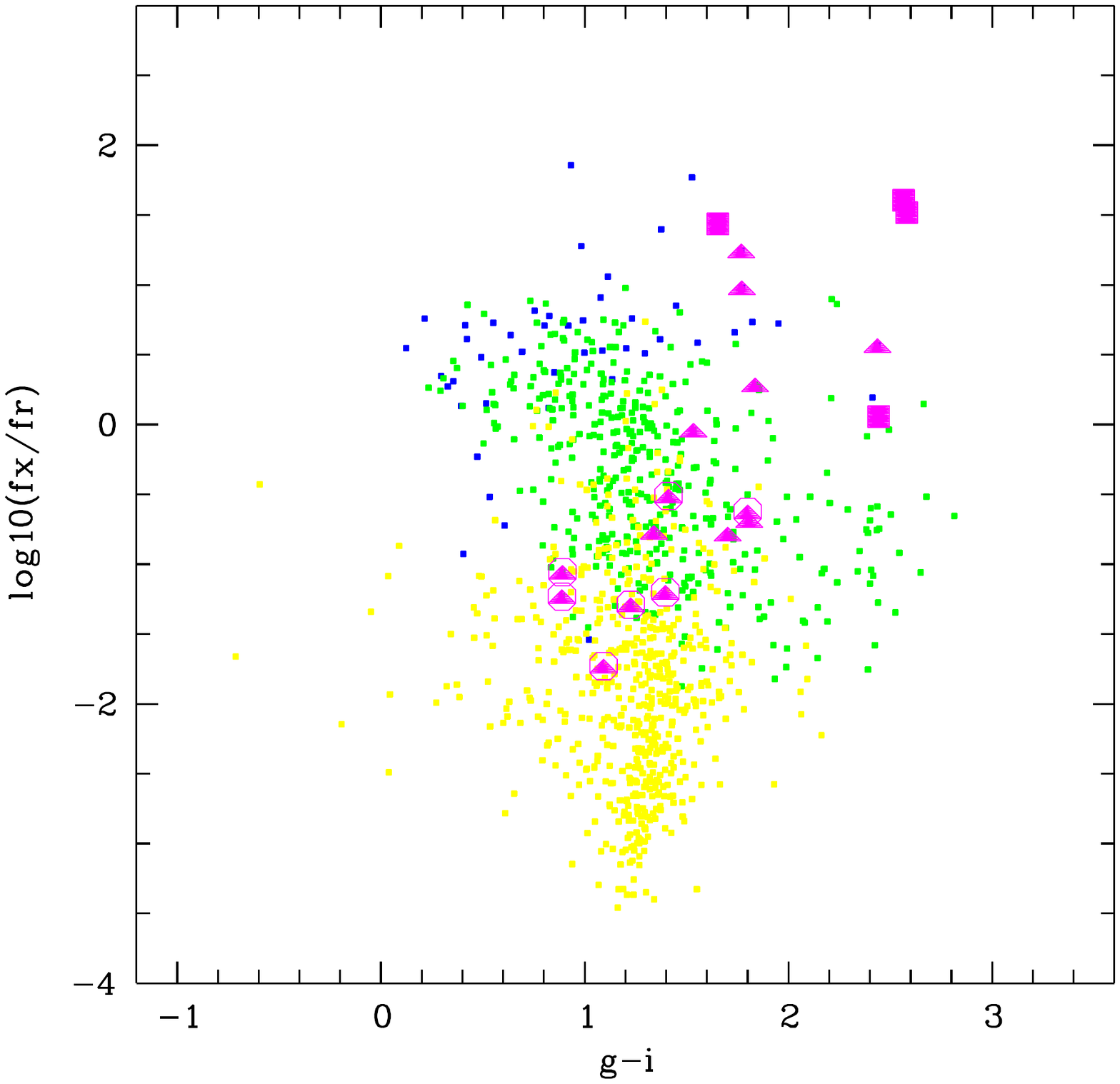} &
      \includegraphics[angle=-90,width=0.5\textwidth,viewport=0 50 550 650]{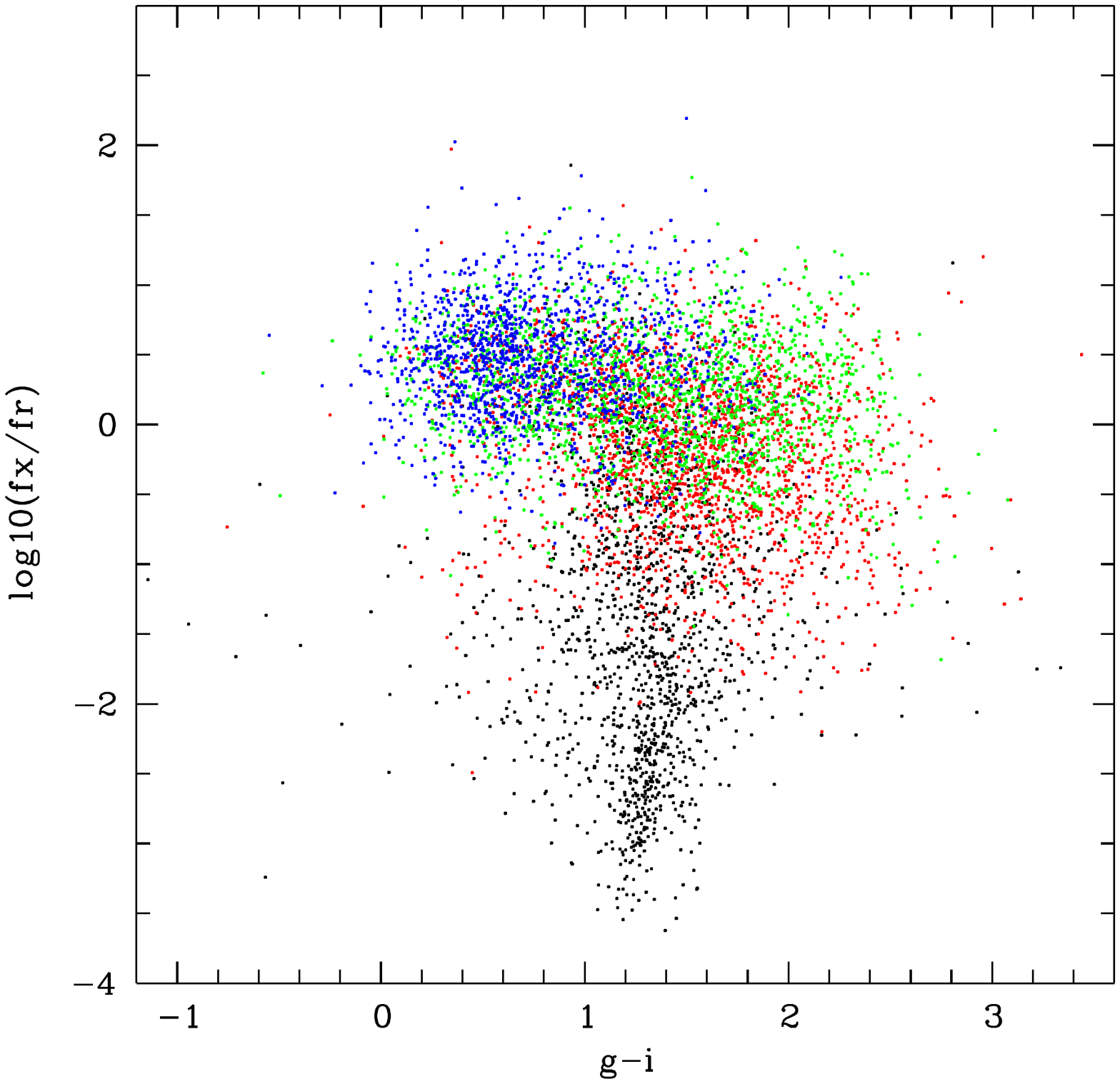} \\
    \end{tabular}        
    \caption{Distribution of spatially resolved objects in the optical band in the $log(f_x/f_r)$ versus $g-i$ diagram. Left: the identified sample blue: AGNs with $log($\Lx$) \geq$ 44, green: $44 \geq log($\Lx$) \geq 42$, yellow: galaxies with $log($\Lx$) \leq$ 42 AGN, magenta: QSO2 - filled squares = X-ray selected - filled triangles = optically selected - encircled = Compton Thick (see \S\ref{qso2}), red: stars, cyan: accreting binaries.  Right: Entire SDSS photometric sample. In this case, the colour codes the range of $r$ magnitude. Black: $<$18, red:18-20, green, 20-21, blue:$>$21. We only show SDSS entries with a probability of identification with an X-ray source larger than 90\%, $g$ and $i$ magnitudes brighter than 22.2 and errors on $g-i$ $<$ 0.2.}
\label{fig:fxfo_vs_gi_ext_type}    
\end{figure*}

The huge merit of the dedicated identification programmes such as the ones carried out by the SSC is to extend spectroscopic identifications to very low optical fluxes and thus offer a unique opportunity to unveil the different populations of extragalactic sources which may appear at fainter optical fluxes and in general at higher X-ray to optical flux ratios. The two strategies, wide and shallow on one hand and narrow and deep on the other hand are indeed quite complementary and suited to best characterise and scientifically investigate the entire serendipitous XMM-Newton catalogues. 

\subsection{The main classes of X-ray sources}

We investigate the distribution of the various classes of X-ray sources in the original instrumental parameter space. In principle, we could also have used the parameter space resulting from a Principal Component Analysis (PCA), thus highlighting the most significant (or information-rich) linear combination of physical measurements. \cite{pineau2008} showed that the two first eigenvectors deriving from the PCA analysis of the 2XMMi/SDSS DR7 sample gathering the largest data variance are indeed close to the two main ones used here, namely X-ray to optical flux ratios and optical colours. However, taking into account eigen-axes of higher orders, which include X-ray spectral information in the form of hardness ratios, can slightly improve the separation between different classes of sources \citep{pineau2008}. 

We show in Figs. \ref{fig:fxfo_vs_gi_pl_type} and \ref{fig:fxfo_vs_gi_ext_type} the positions in the $g-i$ / $log(f_x/f_r)$ diagram of the various classes of objects present in the identified sample (left panel) and of all 2XMMi sources with only a cross-identification with a photometric SDSS entry (right panel). Sources spatially unresolved and extended in the optical band are presented separately in the two figures. 
A majority of 2XMMi sources match with SDSS-DR7 photometric entries close to the limiting magnitude of the SDSS survey (mag $\sim$ 22), and only a relatively small fraction is bright enough to have been selected for spectroscopic observations. For instance, over a grand total of 60567 2XMMi detections matching a SDSS DR7 entry, 87\% have an error on their $r$ magnitude below 0.2, but only 12\% are SDSS spectroscopic targets as well. In addition, the repartition of the spectroscopic targets are far from covering uniformly the parameter space spun by the optical counterparts of the serendipituous XMM-Newton sources. 

\subsection{Separating stellar from extragalactic sources}

As expected, the $f_x/f_r$ ratio is a very powerful parameter to separate the late-type stellar X-ray population in which the high-energy emission arises in a magnetic active corona from X-ray luminous sources powered by accretion such as active galactic nuclei or cataclysmic variables. However, the distribution of low $L_X$ galaxies in $log(f_x/f_r)$  clearly overlaps with that of active coronae. Introducing the $g-i$ colour index allows us to separate the bulk of the stars, especially the reddest M stars from most galaxies. Nevertheless, many galaxies, in particular of the early type, exhibit optical energy distributions similar to those of G type stars, show comparable $(f_x/f_r)$ and consequently cannot be easily distinguished from stars in the $g-i$ / $log(f_x/f_r)$ diagram.  Obviously, taking into account the spatial extension of the optical source allows us to efficiently separate them from stars (see Figs. \ref{fig:fxfo_vs_gi_pl_type} and \ref{fig:fxfo_vs_gi_ext_type}).

Interestingly, the reddest point-like optical sources located on the "stellar" branch are also the faintest ones with $r$ magnitudes in the range of 18 to 20 (see Fig. \ref{fig:fxfo_vs_gi_pl_type}, right panel). They also appear to exhibit the highest $f_x/f_r$ ratio. This is consistent with the known increase of the $f_x/f_{opt}$ ratio for M stars compared to that of earlier spectral types \citep[see e.g.][]{vaiana1981}. We note, however, that some high $z$ QSOs have been identified with very red point-like objects of $f_x/f_r$ ratios approaching those of active coronae (see Fig. \ref{fig:fxfo_vs_gi_pl_type}).

Although cataclysmic variables occupy a locus in the $g-i$ / $log(f_x/f_r)$ comparable to that of most quasars, their distribution exhibits a wider spread than that of AGNs. This large scatter can be used to provide a high likelihood identification of their class, at least for part of them. For instance, very blue objects, typically with $g-i$ below $-$0.2 as well as those with extreme $f_x/f_r$, have a high probability of being cataclysmic variables. 

\subsection{Distinguishing between the various classes of extragalactic sources}

Figure \ref{fig:fxfo_vs_gi_pl_type} shows that many of the 2XMMi sources having a counterpart in the DR7 of the SDSS cluster in a rather narrow range of blueish $g-i$ colours in the interval of $-$0.2 to 0.8. They are characterised by a $log(f_x/f_r)$ $\sim$ 0 and appear as point-like sources in the optical. Their positions in this diagram overlaps with that of the vast majority of the spectroscopic SDSS AGN found in our identified sample, which for most of them are UV-excess optically selected quasars. 

\begin{figure}
      \includegraphics[angle=-90,width=0.5\textwidth,viewport=0 50 550 650]{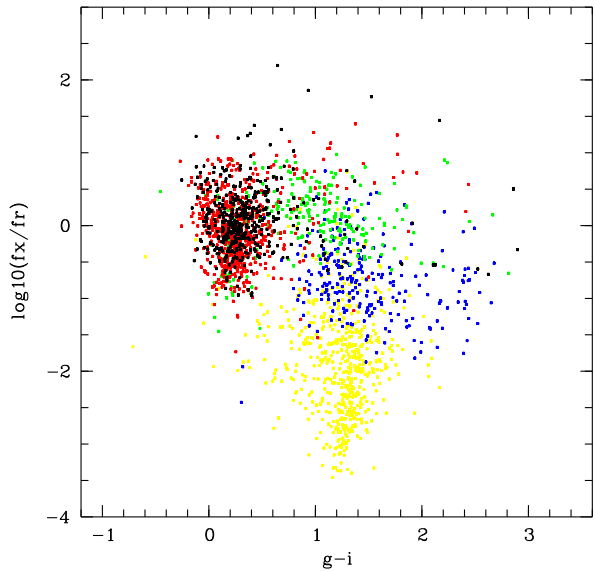}       
      \caption{Distribution of extragalactic objects in the $log(f_x/f_r)$ versus $g-i$ diagram according to \Lx. Black; $>$ 10$^{45}$, red 10$^{44}$ to 10$^{45}$, green 10$^{43}$ to 10$^{44}$, blue 10$^{42}$ to 10$^{43}$,  yellow 10$^{38}$ to 10$^{42}$. We only show SDSS entries with a probability of identification with an X-ray source higher than 90\%, $g$ and $i$ magnitudes brighter than 22.2 and errors on $g-i$ $<$ 0.2.}
      \label{fig:fxfo_vs_gi_lum}
\end{figure}

Let us now consider all objects, both spatially resolved and unresolved, occupying the UV excess quasar region  '$g-i$ values comprised between $-$0.2 and +0.8 and $log(f_x/f_r)$ $\geq$ -1.2). In this range of parameters, the mean $log(f_x/f_r)$ of the spectroscopically identified sample appears slightly shifted by $\sim$ 0.3 dex to lower values (i.e. 0.75 mag brighter for a given X-ray flux), compared to that of the photometric sample. Since the mean $r$ magnitude of the corresponding spectroscopic and photometric-only groups are of 18.86 and 20.73 respectively, as a result of the necessarily brighter optical flux limit of the spectroscopic sample, this indicates that the photometric sample is dominated by a slightly more remote population of AGN, hence fainter in X-rays and in optical than the spectroscopic sample, albeit with a somewhat larger mean $f_x/f_r$. It can also be seen in Fig. \ref{fig:fxfo_vs_gi_lum} that these UVX spectroscopically identified quasars are the most energetic with X-ray luminosities in excess of 10$^{44}$ergs/s. In a general manner, Fig. \ref{fig:fxfo_vs_gi_pl_type} shows that the spectroscopically identified sample of point-like objects covers the range of parameters populated by the photometric cross-identifications for both AGN and stars relatively well, except, as quoted above, for the faintest optical matches. This identified sample could thus be used as a learning sample to statistically identify and classify X-ray sources with optical counterparts of comparable brightness.

This is at variance with the situation prevailing for extended sources. As seen in Fig. \ref{fig:fxfo_vs_gi_ext_type}, a considerable number of X-ray sources are identified with red spatially extended photometric objects, i.e., relatively faint reddish galaxies with $g-i$ $\ga$ 1.0 and corresponding $log(f_x/f_r)$ $\geq$ $-$0.5. Unfortunately, the SDSS policy for selecting spectroscopic targets does not cover this region of the parameter space well. In the few cases in which an optical spectrum exists, they are assigned an AGN type. These galaxies are significantly optically brighter than most UVX quasars, most of which are in the $r$ mag range of 18 to 21. Their derived X-ray luminosities in the range of 10$^{43-44}$\,\ergs\ (Fig. \ref{fig:fxfo_vs_gi_lum}) clearly show that the vast majority of these reddish objects are likely Seyfert galaxies. This population extends downwards to lower $f_x/f_r$ ratios, narrowing the $g-i$ range spanned and decreasing their X-ray luminosities. Eventually the brightest objects ($r$ $\la$ 18) merge with the group of "normal" galaxies with \Lx $\la$ 10$^{42}$\,\ergs, which is well represented in the spectroscopically identified sample. These low X-ray luminosities could be explained in terms of ULXs, starbursts, or of a collection of low-mass X-ray binaries in elliptical galaxies. 

\begin{table*}
\centering
\caption{Main properties of the 2XMMi/SDSS identifications with $log(f_x/f_r) \geq -0.5$ in each 0.5 $g-i$ bin.}  
\begin{tabular}{cccccc}
\hline \hline
  $g-i$  &  Ntot	 &  $r$ mag (rms) &   pn HR2 (rms) & pn HR3 (rms) & pn HR4 (rms) \\
\hline
\multicolumn{6}{c}{point-like sources} \\ \hline
 +0.25     &  9538&+20.20 (1.07) & -0.115 (0.228) & -0.397 (0.237) & -0.219 (0.376) \\
 +0.75     &  3034&+20.59 (0.99) & -0.045 (0.250) & -0.360 (0.246) & -0.199 (0.330) \\
 +1.25     &   471&+20.51 (0.84) & +0.061 (0.352) & -0.294 (0.317) & -0.188 (0.338) \\
 +1.75     &	96&+20.17 (0.97) & +0.168 (0.280) & -0.229 (0.294) & -0.150 (0.347) \\
 +2.25     &	31&+19.79 (1.02) & +0.200 (0.315) & -0.307 (0.282) & -0.168 (0.598) \\
 +2.75     &	29&+19.04 (1.31) & -0.081 (0.368) & -0.447 (0.302) & -0.162 (0.545) \\
\hline 
\multicolumn{6}{c}{extended sources} \\ \hline
 +0.25     &   812&+20.97 (0.84) & -0.099 (0.222) & -0.385 (0.255) & -0.175 (0.425) \\
 +0.75     &  1915&+20.23 (1.46) & -0.100 (0.237) & -0.370 (0.235) & -0.246 (0.311) \\
 +1.25     &  1918&+19.77 (1.44) & -0.009 (0.307) & -0.285 (0.335) & -0.149 (0.341) \\
 +1.75     &  1364&+19.89 (0.97) & +0.172 (0.377) & -0.111 (0.452) & -0.028 (0.383) \\
 +2.25     &   479&+20.00 (0.60) & +0.233 (0.414) & -0.012 (0.482) & -0.018 (0.337) \\
 +2.75     &    35&+19.81 (0.78) & +0.027 (0.456) & +0.056 (0.499) & +0.027 (0.395) \\
\hline
\end{tabular}
\note{We only consider here sources with error on $g-i$ $\leq$ 0.2, $g$ and $i$ mags brighter than 22.2 and individual probability of identification greater than 90\%.}
\label{redids}
\end{table*}

\section{Science cases}

In the next two sections we touch upon two distinct science cases, one in the extragalactic domain and one related to a galactic source population.  These two examples aim at illustrating the range of research that these clean cross-correlated samples allow and do not explore in depth all possible paths of investigations. In particular, we do not make use of the spectroscopic line data, which could provide many additional astrophysical diagnostics. The first case considered bears on the topical search for
QSO2s, while the second one explores the X-ray and optical properties of active stellar coronae.  

\subsection{Searching for QSO2 candidates}\label{qso2}

\begin{figure*}
    \begin{tabular}{cc}
      \includegraphics[angle=-90,width=0.5\textwidth,viewport=0 50 550 650]{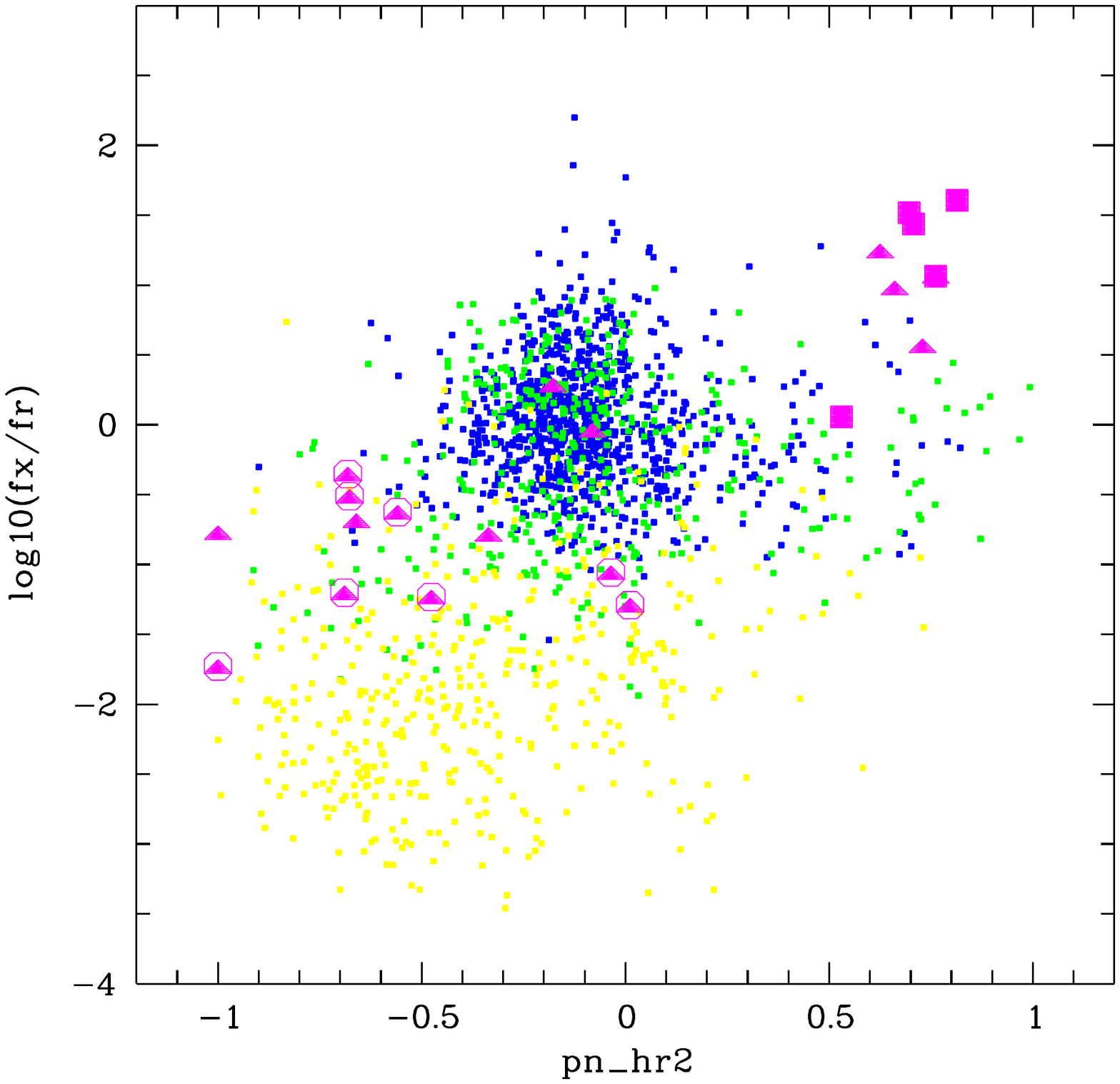} &
      \includegraphics[angle=-90,width=0.5\textwidth,viewport=0 50 550 650]{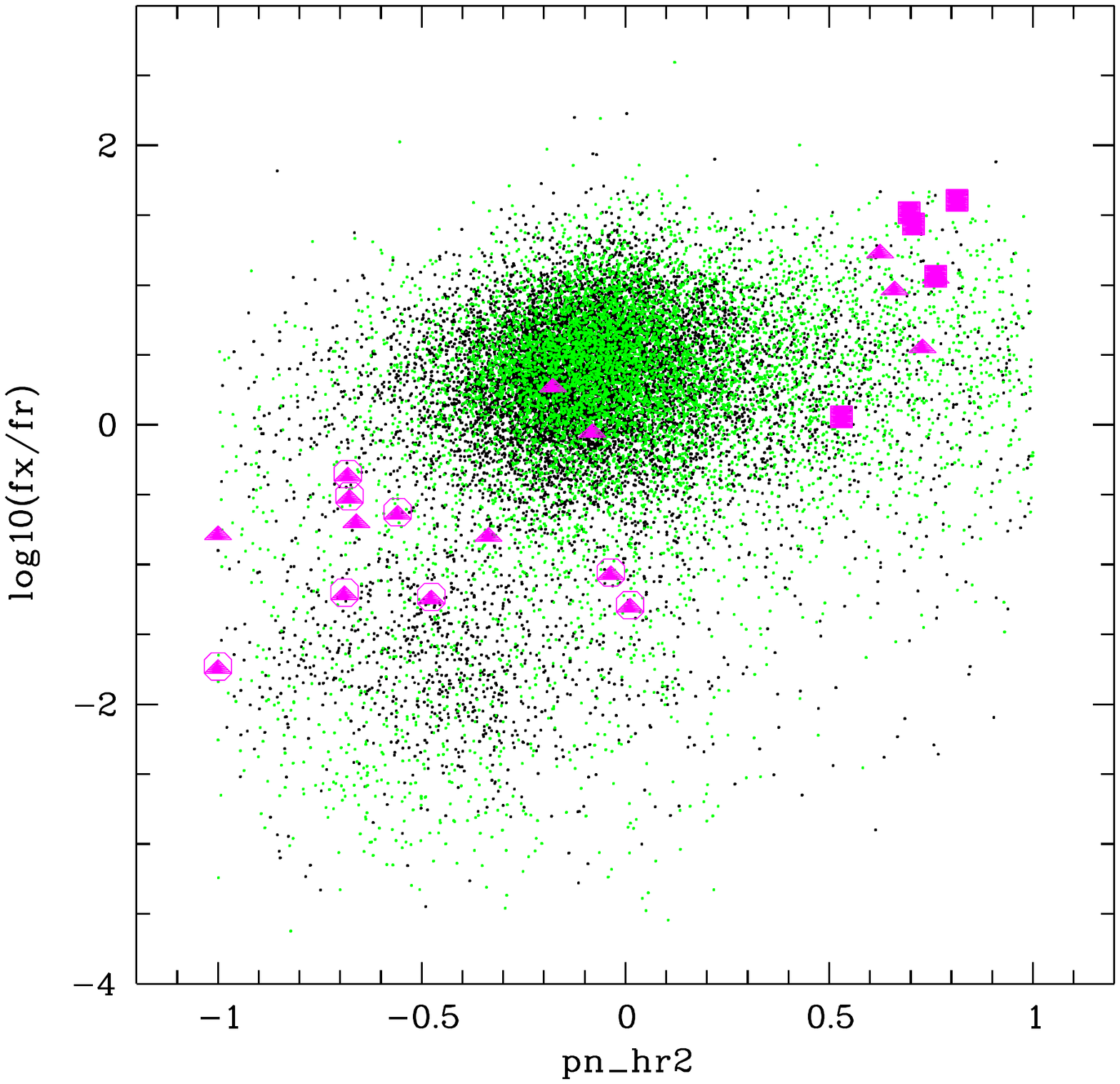} \\
    \end{tabular}        
    \caption{   
Left panel: Distribution of spectroscopically identified objects (optically resolved and unresolved) in the $log(f_x/f_r)$ versus EPIC pn HR2 diagram. blue: AGNs with $log($\Lx$) \geq$ 44, green: $44 \geq log($\Lx$) \geq 42$, yellow: galaxies with $log($\Lx$) \leq$ 42, magenta: Type 2 QSO - filled squares = X-ray selected - filled triangles = optically selected - encircled = Compton Thick. Right panel: the entire SDSS photometric sample.  Black: unresolved objects, green: extended objects, magenta: same as in right panel. 
We only show SDSS entries with a probability of identification with an X-ray source higher than 90\%, $r$ magnitudes brighter than 23 and errors on HR2 less than 0.3.}
\label{QSO2_xo_hr2}    
\end{figure*}

\begin{figure*}
    \begin{tabular}{cc}
      \includegraphics[angle=-90,width=0.5\textwidth,viewport=0 50 550 650]{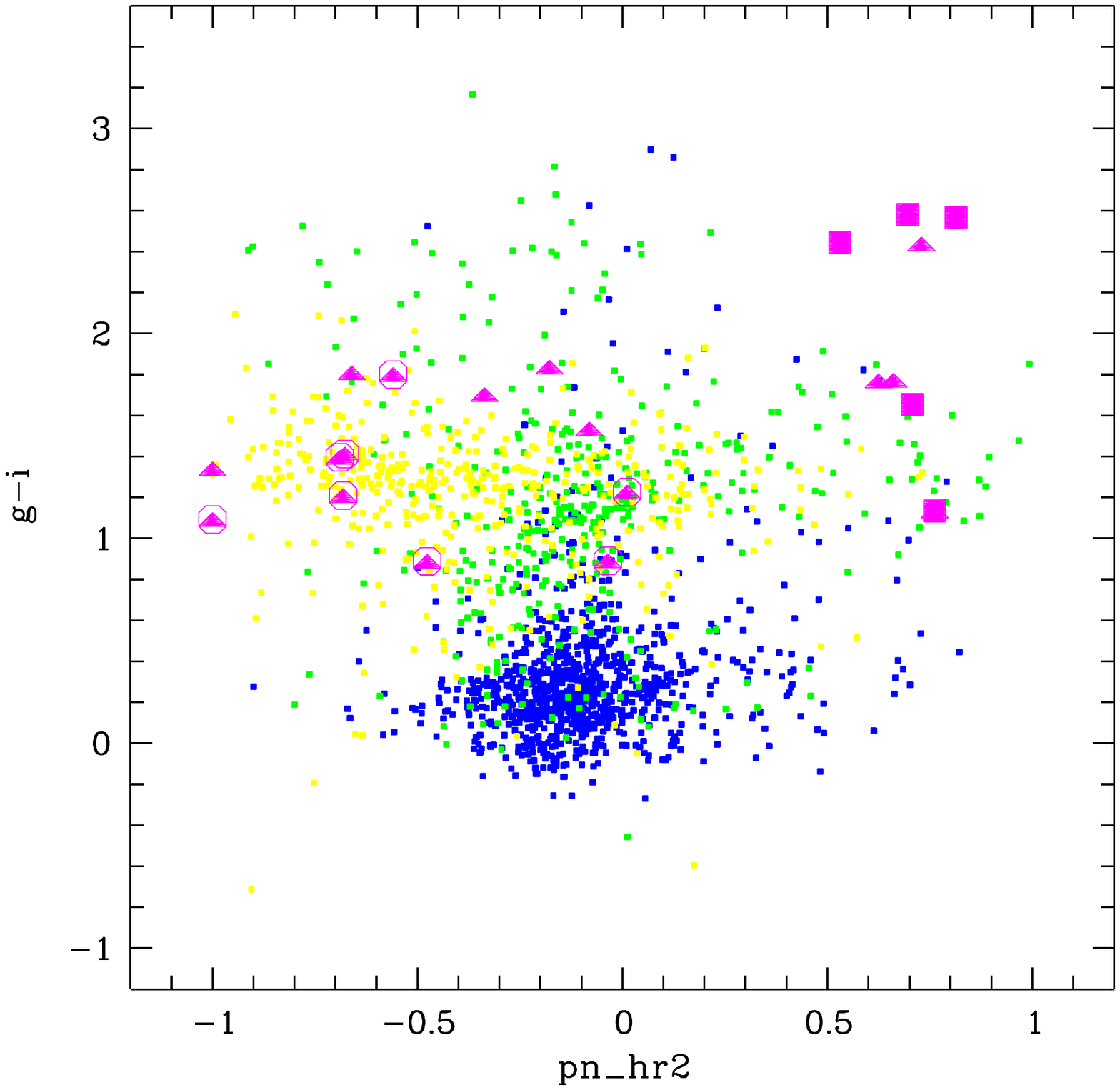} &
      \includegraphics[angle=-90,width=0.5\textwidth,viewport=0 50 550 650]{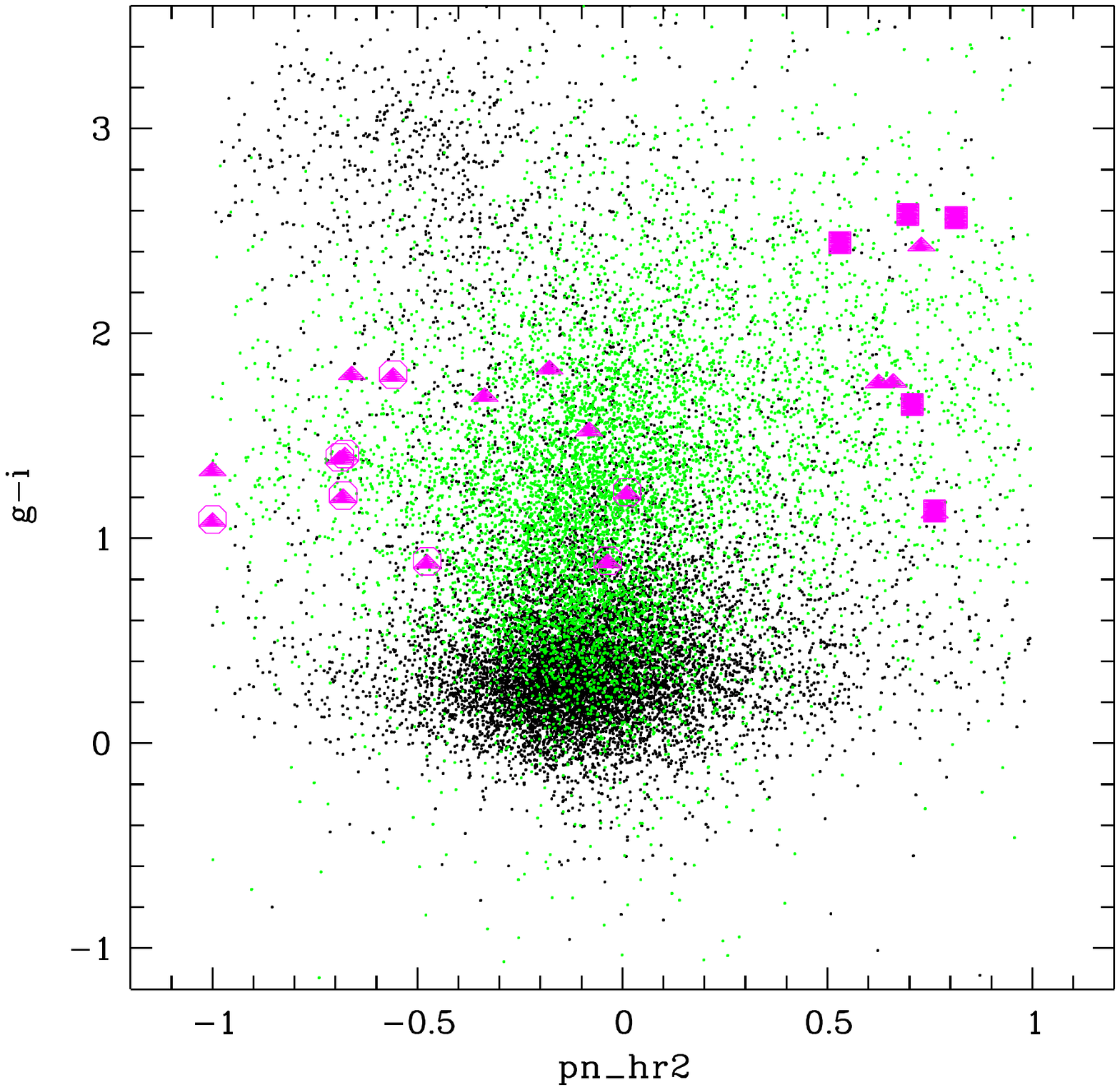} \\
    \end{tabular}        
    \caption{Left panel: Distribution of spectroscopically identified objects (optically resolved and unresolved) in the $g-i$ versus EPIC pn HR2 diagram. blue: AGNs with $log($\Lx$) \geq$ 44, green: $44 \geq log($\Lx$) \geq 42$, yellow: galaxies with $log($\Lx$) \leq$ 42, magenta: Type 2 QSO - filled squares = X-ray selected - filled triangles = optically selected - encircled = Compton Thick. Right panel: the entire SDSS photometric sample.  Black: unresolved objects, green: extended objects, magenta: same as in right panel. We only show SDSS entries with a probability of identification with an X-ray source larger than 90\%, $r$ magnitudes brighter than 23 and errors on HR2 less than 0.3.} 
\label{QSO2_gi_hr2}    
\end{figure*}

The members of the high-luminosity high-obscuration part of the AGN
population are commonly denominated QSO2s. The synthesis modelling of the XRB
\citep{gilli2007,treister2009} predict that up to $\sim 20\%$ to the
XRB \citep{gilli2007} could be produced by QSO2s, they could represent
$\sim 30-50\%$ of the high luminosity AGN population
\citep[e.g.][]{dellaceca2008}, and they could probably co-evolve with
massive host galaxies \citep{severgnini2006}; it is therefore clear, that
hunting for QSO2 remains one of the most topical activities as is the
search for associated X-ray and optical signatures.

Optical selection of QSO2s
relies on finding objects showing only narrow emission lines with
high-ionisation line ratios and high luminosity typically from
[OIII], e.g. $L_{[O{\sc III}]} > 10^{8.3} L_{\odot}$
\citep{zakamska2003,reyes2008}. X-ray selection looks instead for luminous
($L_{X,2-10{\mathrm keV}}>10^{44}$~erg/s) significantly obscured
(column density $N_H>10^{22}\,{\mathrm cm}^{-2}$) sources, which are
best selected in the $E>2$~keV hard X--ray band
\citep[e.g.][]{mainieri2002,caccianiga2004,perola2004,vignali2006,dellaceca2008,krumpe2008}. Within
the Unified Model \citep{antonucci1993}, obscuration of the central
X-ray-emitting and Broad Line-emitting regions by an intervening torus
gives rise to those consistent properties across both bands.

A priori, QSO2s should present red optical colours (since the emission
of the host galaxy would dominate over the obscured AGN), high X-ray
hardness ratios\footnote{The XMM-Newton EPIC
  hardness ratios use five energy bands, which expressed in keV units
  are 0.2--0.5, 0.5--1, 1--2, 2--4.5 and 4.5--12.0. Hardness ratio $i$
  is expressed as
\begin{equation}
\rm HR_{i} = \frac{C_{i+1} - C_{i}}{C_{i+1} + C_{i}} \ ,
\end{equation} 
with $C_{i}$ the count rate in band {i} corrected for vignetting.}
(because of the predominant absorption of the lower
energy X-rays) and high X-ray-to-optical flux ratio ($f_x/f_r$, since the X-rays
are less sensitive to absorption than the optical range). However,
several effects could alter this simple recipe.  For instance, Compton
Thick absorption (defined here as \nh $\ga$ 10$^{24}$cm\,$^{-2}$)
would completely absorb direct X-rays up to 10~keV, this would alter
both the spectral shape (since scattered primary X-rays have
``softer'' spectra) and the ratio of optical to X-ray fluxes would be
more typical of ``normal'' galaxies (i.e. $\log(f_x/f_r)<-1$). 

We investigate here whether the position in the overall X-ray and
optical parameter space of the confirmed/candidates QSO2 discovered so
far could give some hint on the way other candidates could be selected
on the basis of broadband high-energy XMM-Newton data and optical
photometry only.

To do so we first assembled a sample of {\it bona fide} QSO2 from
optical and X-ray surveys. The optically selected sample has been
obtained by cross-correlating the SDSS sample of \cite{zakamska2003}
and \cite{reyes2008} with the 2XMMi catalogue. We list in Table
\ref{zakamska} the main properties of the QSO2 SDSS candidates
matching a serendipitous EPIC source. The listed X-ray luminosities
were computed assuming an average shape (see Sect. 
\ref{finalidsamp}) for the large band 0.2 to 12\,keV energy
distribution and are not corrected for intrinsic 
absorption\footnote{These luminosities are necessarily less accurate than those derived by
\cite{ptak2006} from a detailed spectral analysis; however the four
candidates in the list of \cite{zakamska2003} in common with the eight
discussed in \cite{ptak2006} (all Compton Thin QSO2) have
luminosities consistent within a factor of two.}.  
We also marked in the table the eight SDSS QSO2 that are good candidates to be Compton
Thick.  To define these objects we used $L_{X,meas}/L_{X,O[III]}
< 0.05$\footnote{Our criteria to define a Compton Thick AGN is slightly less
  restrictive than that recently used by e.g. \cite{vignali2010},
  $L_{X,meas}/L_{X,O[III]} < \sim 0.01$.}, where L$_{X,meas}$ is the
observed 2-10 keV luminosity (see above) and L$_{X,[OIII]}$ is the
expected intrinsic X-ray luminosity (the latter has been computed
using the observed $L_{O[III]}$ and the ratio $L_{O[III]}/L_{x} \sim
0.017$ derived for the unobscured view of Seyfert galaxies
\citep{heckman2005}.).

To this sample of optically selected QSO2 we added a small sample
of five X-ray selected QSO2 (with the definition above) obtained by
cross-correlating SDSS, 2XMMi, and a few selected lists of X-ray
defined QSO2 \citep{dellaceca2008,krumpe2008,corral2010}. 
It is worth stressing that based on a detailed
analysis of the X-ray and optical spectral properties
\citep[see][]{dellaceca2008,corral2010}, all these X-ray defined QSO2
are Compton Thin (intrinsic $N_H$ between $10^{22}$ cm$^{-2}$ and few
times $10^{23}$ cm$^{-2}$).

\begin{table*}
    \centering
    \caption{QSO2 candidates from \cite{zakamska2003}, \cite{reyes2008} and X-ray selected samples matching 2XMMi entries with EPIC pn observations.}
    \begin{tabular}{lccccrcr}
\hline
2XMMi name		& id prob & z  &  u-g          &     g-i	& HR 2 (pn) &  Log(\Lx) & Log($f_x/f_r$) \\ 
\hline
                         \multicolumn{7}{c}{Zakamska et al. (2003)} \\
\hline 
 2XMM J005621.6+003235$^{Tck}$ &0.964&0.484&1.891 $\pm$0.643&1.800 $\pm$0.061&-0.56 $\pm$0.28&42.78 & -0.626\\ 
 2XMMi J011522.2+00151         &0.997&0.390&6.023 $\pm$1.239&2.437 $\pm$0.046& 0.73 $\pm$0.04&44.26 &  0.566\\ 
 2XMM J015716.9-005305         &0.988&0.540&0.565 $\pm$0.167&1.836 $\pm$0.050&-0.18 $\pm$0.46&42.78 &  0.285\\ 
 2XMM J021047.0-100152         &0.999&0.540&0.348 $\pm$0.136&1.771 $\pm$0.047& 0.66 $\pm$0.12&44.40 &  0.980\\ 
 2XMM J103951.5+643005$^{Tck}$ &0.991&0.402&0.490 $\pm$0.075&1.209 $\pm$0.027&-0.68 $\pm$0.33&43.02 & -0.353\\ 
 2XMM J122656.4+013124         &0.998&0.732&0.970 $\pm$0.256&1.134 $\pm$0.056& 0.76 $\pm$0.06&44.83 &  1.063\\ 
 2XMM J164131.6+385841         &0.995&0.596&0.550 $\pm$0.112&1.768 $\pm$0.028& 0.63 $\pm$0.05&44.96 &  1.245\\ 
\hline
\multicolumn{7}{c}{Additional candidates from Reyes et al. (2008)} \\
\hline
 2XMMi J075821.2+392337$^{Tck}$ &0.912 &0.216 &0.620 $\pm$0.026 &0.891 $\pm$0.010  &-0.04 $\pm$0.27& 42.37&-1.058\\
 2XMM J093952.7+355358          &1.000 &0.137 &1.527 $\pm$0.051 &1.534 $\pm$0.007  &-0.08 $\pm$0.06& 43.44&-0.040\\
 2XMM J094506.4+035552$^{Tck}$  &0.995 &0.156 &0.979 $\pm$0.043 &1.092 $\pm$0.025  &-1.00 $\pm$0.28& 41.63&-1.730\\
 2XMM J100327.8+554155          &0.991 &0.146 &1.390 $\pm$0.066 &1.340 $\pm$0.010  &-1.00 $\pm$0.17& 42.40&-0.772\\
 2XMM J103408.5+600152$^{Tck}$  &1.000 &0.051 &1.408 $\pm$0.009 &0.888 $\pm$0.003  &-0.48 $\pm$0.05& 42.29&-1.232\\
 2XMM J103456.3+393939$^{Tck}$  &0.995 &0.151 &1.465 $\pm$0.049 &1.412 $\pm$0.007  &-0.68 $\pm$0.29& 42.88&-0.512\\
 2XMM J122709.8+124855          &0.998 &0.194 &1.728 $\pm$0.162 &1.808 $\pm$0.010  &-0.66 $\pm$0.27& 42.88&-0.686\\
 2XMM J131104.6+272806          &0.998 &0.240 &1.272 $\pm$0.070 &1.702 $\pm$0.010  &-0.34 $\pm$0.07& 42.94&-0.784\\
 2XMM J132419.8+053704$^{Tck}$  &0.997 &0.203 &0.441 $\pm$0.027 &1.226 $\pm$0.009  & 0.01 $\pm$0.47& 42.24&-1.290\\
 2XMM J171350.7+572955$^{Tck}$  &0.883 &0.113 &1.279 $\pm$0.025 &1.396 $\pm$0.006  &-0.69 $\pm$0.21& 42.10&-1.201\\
\hline
\multicolumn{7}{c}{X-ray selected QSO2} \\
\hline
 2XMM J113148.6+311400          & 0.995  &0.50   &0.724 $\pm$0.491   &1.654  $\pm$0.091   &0.71 $\pm$0.16& 44.70& 1.437\\
 2XMM J122656.4+013124          & 0.998  &0.73  &0.970 $\pm$0.256   &1.134  $\pm$0.056   &0.76 $\pm$0.06& 44.83& 1.063\\
 2XMM J134656.6+580316          & 0.965  &0.37  &1.472 $\pm$0.538   &2.443  $\pm$0.069   &0.53 $\pm$0.23& 43.88& 0.057\\
 2XMM J160645.9+081523          & 0.960  &0.62  &0.919 $\pm$1.085   &2.565  $\pm$0.216   &0.81 $\pm$0.26& 44.80& 1.607\\
 2XMM J204043.2$-$004548        & 0.336  &0.62  &1.448 $\pm$2.334   &2.581  $\pm$0.304   &0.70 $\pm$0.11& 44.72& 1.518\\
\hline    
\end{tabular}
    \label{zakamska}
    \note{The source 2XMM J122656.4+013124 is present both in the optically selected and in the X-ray selected QSO2 sample.}
\end{table*}

The position of the confirmed/candidates QSO2 in the parameter space
obtained using $f_x/f_r$, the
optical colours (in particular, $g-i$) and hardness ratio (in
particular HR2) are shown in
Fig.\ref{QSO2_xo_hr2} and in Fig.\ref{QSO2_gi_hr2}; we marked
with different symbols the several ``flavours" of QSO2, i.e. the X-ray
selected QSO2 (all Compton Thin), the optically selected QSO2 and the
candidate Compton Thick QSO2.
 
As can be seen in Fig.\ref{QSO2_gi_hr2}, the QSO2 generally
appear slightly redder than the bulk of the unresolved AGN, and of the
identified galaxy sample having \Lx\,$>\, 10^{44}$ \ergs.  
Their colours are more similar to that of the
galaxy identified sample with $10^{42} < $\Lx$ < 10^{44}$
\ergs\ (green points in Fig.\ref{QSO2_gi_hr2}, left panel).

The position of the QSO2 in Fig.\ref{QSO2_xo_hr2} and in
Fig.\ref{QSO2_gi_hr2} clearly shows a separation between the
``confirmed" Compton Thin QSO2 (magenta filled squares
and triangles) and the ``candidates" Compton Thick QSO2
(encircled). Interestingly, the four ``optically selected" QSO2
occupying the same region of the X-ray selected Compton Thin QSO2 have
been studied in the X-ray domain by \citep{ptak2006}; all these
sources (2XMMiJ011522.2+001518, 2XMMiJ021047.0$-$100152,
2XMMiJ122656.4+013124, 2XMMiJ164131.6+385841) are described by an
absorbed power-law model with an intrinsic $N_H \sim 2-3 \times
10^{22}$ cm$^{-2}$. Therefore the upper right corner of
Fig.\ref{QSO2_xo_hr2} is probably the best
place where to look for Compton Thin QSO2; as shown in
\cite[e.g.][]{caccianiga2004} the very positive HR2 reflects the
relatively large intrinsic photoelectric absorption present in many
QSO2 and responsible for their preferential discovery in hard X-ray
surveys\footnote{We caution however that the separation in HR2
properties between Compton Thin and Compton Thick QSO2 is probably 
clear only for sources with redshift as high as in our learning
sample ($z < \sim$ 0.8); at higher redshift we should measure
decreasing HR2 values for Compton Thin QSO2 because the observed X-ray
energy band will move to higher energies, which are increasingly less
affected by absorption.}.

\begin{figure}
    \centering
    \includegraphics[angle=-90,width=0.5\textwidth,viewport=0 50 550 650]{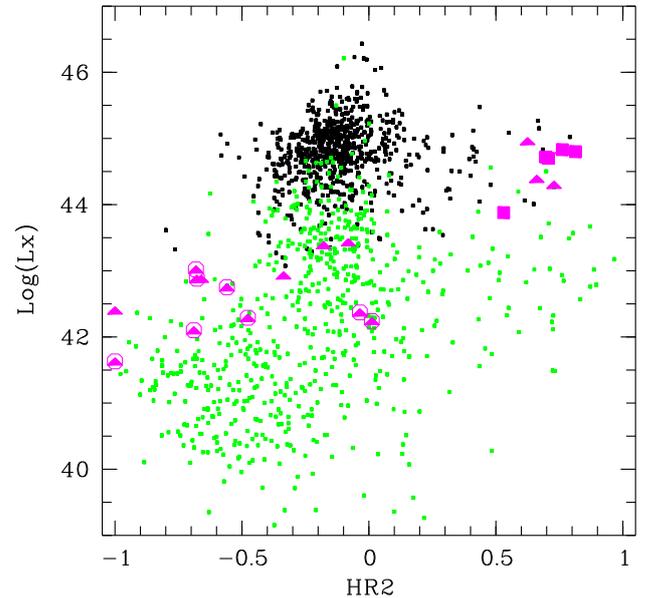} 
    \caption{Variation of the EPIC pn hardness ratios HR2 with X-ray luminosity for extragalactic spectroscopic SDSS targets optically unresolved (black dots) and optically extended (green dots). Magenta: Type 2 QSO - filled squares = X-ray selected - filled triangles = optically selected - encircled = Compton Thick. Errors on hardness ratios are below 0.2 in all cases.}
\label{plot_agn_Lx_Hr}
\end{figure}

Finally, we show in Fig. \ref{plot_agn_Lx_Hr} the behaviour of the
EPIC pn HR2 hardness ratio with X-ray luminosity for all spectroscopic
SDSS targets. The bulk of the SDSS Type 1 QSOs with X-ray luminosities
higher than 10$^{44}$\ergs\ (0.2-12 keV) cluster around an hardness
ratio $<$HR2$>$ = $-$0.12; the same objects cluster around hardness
ratios $<$HR3$>$ = $-$0.38 and $<$HR4$>$ = $-$0.28. These hardness
ratios are in excellent agreement with the values expected from a
canonical $\Gamma$ = 1.9 power law X-ray spectrum undergoing
negligible intrinsic absorption and a mean Galactic absorption of 1.16
$\times$ 10$^{20}$cm$^{-2}$ (the average over all directions of galaxy
and QSO targets). For this group of QSOs, there is no evidence of
strong dependence of the power-law index with \Lx. There is however a
small number of QSOs exhibiting a considerably harder X-ray spectrum
(as testified by an increasing value of HR2) extending to the same
locus occupied by confirmed Compton Thin QSO2. As one enters the AGN regime 
at X-ray luminosities below
10$^{44}$\ergs, the number of extended sources with galaxy-like
optical spectra rises considerably and the shape of the X-ray energy
distribution shows a much larger scatter.  The candidates Compton
Thick QSO2 seem to populate this part of the diagram. 
 
A number of spectroscopic SDSS entries occupy the same
region of the LX / HR2 diagram as the reference Compton Thin
QSO2. We explored the nature of these candidates by selecting
objects with log(\Lx) higher than 44, EPIC pn HR2 greater
than 0.5 with an error of less than 0.2 on the hardness ratio:
twelve objects match these conditions.
From an inspection of the spectroscopic SDSS data and a literature search we found
that at least $\sim$ 60\% of them are indeed characterised  by absorption at same
level: two objects are clearly Broad Absorption Line QSOs (SDSS
J114312.32+200346.0, SDSS J141546.24+112943.4), two are  ``dust reddened QSOs''
(SDSS J122637.02+013016.0, SDSS J143513.90+484149.2)  and three are Type 2 QSOs
(SDSS J105144.24+353930.7, SDSS J130005.34+163214.8,  SDSS J134507.93$-$001900.9).
The remaining five objects are apparently ``normal" type I AGN, without any specific
comment in literature: a detailed analysis  of their optical and X-ray properties
(e.g. to understand if these  latter objects could also be classified as ``dust
reddened QSOs") is beyond the  scope of the present paper. The main results of this
exploration is that, although Compton Thin QSO2 do separate well in the
LX / HR2 diagram, other rare kinds of objects occupy the same
locus.

\begin{figure}
    \centering
    \includegraphics[angle=-90,width=0.45\textwidth,viewport=40 33 550 780]{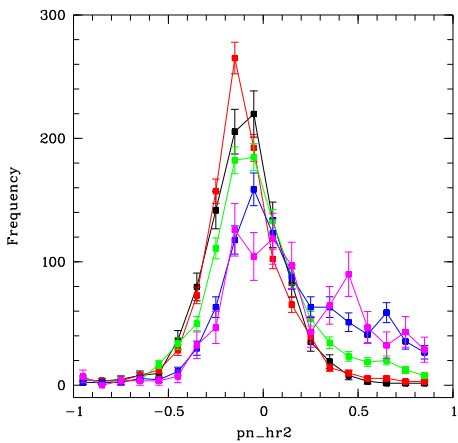}
   \caption{Variation of the EPIC pn hardness ratio HR2 with $g-i$ colour for all 2XMMi matches with extended sources in the entire SDSS catalogue only having a $log(f_x/f_r)$ larger than $-$0.5. Each histogram corresponds to a $g-i$ colour bin of 0.5. Black; $g-i$=0.25, red; 0.75, green; 1.25, blue; 1.75, magenta; 2.25.  Errors on hardness ratios are below 0.3 in all cases. We only show identifications with an individual identification probability higher than 90\%, $g$ and $i$ mag brighter than 22.2, errors on $g-i$ less than 0.2. Histograms are normalised to the total number of sources in the colour bins shown in Table \ref{redids}.}
    \label{histo_gi_ph_mag}
\end{figure}

Finally we investigated the possible presence of Compton Thin
QSO2 candidates in the SDSS photometric matches.  We did this by
looking for sources showing evidences of photoelectric absorption
among the 2XMMi sources matching mainly optically extended SDSS
objects characterised by a high $f_x/f_r$ ratio and relatively red
$g-i$ colours. We thus built histograms of the EPIC pn hardness ratios
for all 2XMMi/SDSS matches with $log(f_x/f_r) \, \ga\, -0.5$ for
various $g-i$ colour intervals. The
resulting histogram for the EPIC pn HR2 hardness ratio is shown in
Fig. \ref{histo_gi_ph_mag}.

The histograms clearly show a main peak of hardness ratio
corresponding to the canonical low \nh \ $\Gamma$ = 1.9 power-law
spectrum characteristic of the type I QSOs for all $g-i$ colour
indexes. However, for the reddest objects, typically with $g-i$ $\ga$
1.5, a secondary bump is observed for harder hardness ratios
(HR2$\simeq$ 0.5) with values consistent with those of the Compton
Thin QSO2; a similar secondary bump is also present if we use the
hardness ratios HR3 (HR4$\simeq$ 0.35) and HR4 (HR4$\simeq$ 0.15). A
detailed investigation of these possibly interesting sources is beyond
the scope of the present paper.

\subsection{X-ray active stars}

At the high galactic latitudes covered here by the legacy SDSS survey ($| b| \ \ga 20$\degr\ with a mean $|b|$ of 58\degr), most of the X-ray sources are of extragalactic origin. For instance, the Extended Medium Sensitivity Survey, which constitutes the largest optically identified sample of serendipitous Einstein X-ray sources at high galactic latitude, contains only 25\% of active stars \citep{stocke1991}. Optical identification campains of the ROSAT all-sky survey sources \citep[RASS, see ][for the bright sample]{schwope2000} yielded similar results (e.g. 35\% of active stars in \cite{Zickgraf1997}), while the ROSAT deep survey of the Lockman field \citep{schmidt1998}, which is $\sim$ 50 times more sensitive that the RASS, collected less than 10\% of stellar sources. Stars constitute a bounded population of comparatively soft X-ray sources. Consequently, their relative contributions to the high galactic source number count is expected to decrease very significantly with increasing sensitivity in Chandra and XMM-Newton observations, which both offer a lower flux limit and harder energy response (see for instance Fig. 1 in \cite{herent2006}). Active coronae indeed constitute less than 10\% of all sources identified in the XMM-Newton serendipitous survey of \cite{barcons2007}. The deepest Chandra surveys such as the Chandra Deep Field North \citep{alexander2003} counts only $\sim$ 3\% of stars.

\begin{figure}
    \centering
    \includegraphics[angle=-90,width=0.5\textwidth,viewport=40 33 550 780]{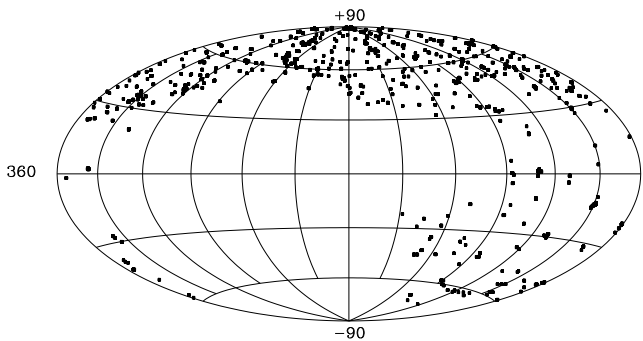}
    \caption{Aitoff projection of the distribution of the 2XMMi sources matching identified stars with a probability $\geq$ 90\%.}   
\label{stars_aitoff}
\end{figure}

This is at variance with the low latitude situation where most of the RASS X-ray sources were found to be associated with stars \citep{motch1997}. XMM-Newton and Chandra have extended to lower fluxes the predominance of active coronae on the soft X-ray low latitude source population \citep{motch2003,rogel2006}. At higher energies, a longitude dependent population of galactic hard X-ray sources appears on the top of a usually dominant background of extragalactic sources \citep{hands2004,motch2006,motch2010}.   

As stated above, the number of X-ray emitting and spectroscopically identified stars available in the DR7 is relatively small, and because of the scientific goals put forward at the time of the selection of targets for spectroscopic follow-up, concentrates on the reddest M type stars. The availability of the SEGUE archive in DR7 has somewhat increased the number of spectroscopically identified stars, but its effect on the cross-correlation statistics remains small. As mentioned in Sect. \ref{buildingls}, in order to increase the stellar sample towards earlier types, we used a kernel density classification to identify the SDSS DR7 / 2XMMi matches with multicolour properties consistent with those expected from stars of main sequence class. 

Figure \ref{stars_aitoff} shows the distribution in galactic coordinates of all X-ray active stars present in the identified sample. Including SEGUE data in the DR7 has allowed the identification of a few stars at low galactic latitude. However, the mean $|b|$ of the stellar sample remains high, ($|b|$ $\sim$ 45\degr) and therefore does not change the conclusion that the present stellar sample is typical of the high Galactic latitudes.

At high $b$  and typical distances of a few hundred parsecs, interstellar absorption remains negligible compared to other uncertainties and has little effect on the observed stellar colours. We computed the total galactic reddening in the directions of each of the X-ray emitting stars following \cite{schlegel1998}. The average E(B-V) is 0.029 with a rms of 0.025. Assuming the absorption coefficients computed by \citep{girardi2004} (\Teff = 4500K; \logg = 4.5), the maximum reddening applicable on average to our sample of stars would be of 0.028 and 0.018 in the $g-r$ and $r-i$ colours respectively.  
Only very few spatially unresolved SDSS sources matching 2XMMi entries have combined $u-g$, $g-r$ and $r-i$ colours compatible with those expected from giant class III stars, not to mention supergiants. In a recent paper, \cite{guillout2009} estimate that their sample of $|b| \la$ 30\degr\ RASS sources identified with bright Tycho stars has a mean contamination of 35\% by evolved stars with a peak at 60\% for K stars. However, a smaller fraction of X-ray emitting evolved stars of $\sim$ 10\% was present in the sample of \cite{covey2008}, which is more representative of what we should expect in our case because it was selected at higher galactic latitude and fainter flux. Below, we will assume that all stars belong to the main sequence, keeping in mind that a fraction of the class III and class IV stars, in particular in short period binaries such as RS CVn systems, could contribute to some extent. Being considered as single dwarfs, these stars would have computed X-ray luminosities below that actually emitted. 

Our sample comprises 549 active coronae candidates with individual probability of identification above 90\%, corresponding to a total sample reliability of 98\%. The interval of $g-r$ and $g-i$ colours corresponds to K4 to late M5-M6 stars. Earlier stars are more often subject to optical saturation and have in general lower KDC probabilities. They are therefore excluded from the clean identified stellar sample built here.

Neglecting reddening effects, we computed the distances using the absolute magnitude calibration listed in \cite{covey2007} and X-ray luminosity using the broad band (0.2-12 keV) flux listed in the 2XMMi catalogue for the EPIC camera.  The mean photometric distance is 340\,pc and all stars have distances in the range of 40\,pc to 2000\,pc. The overall Log(Lx) (0.2--12\,keV) distribution peaks at 29.20 and ranges from $\sim$ 27.1 to $\sim$ 30.9. This interval of X-ray luminosity covers that exhibited by old stars such as the Sun or even less active, up to that emitted by the most active T Tauri or RS CVn stars. The mean X-ray luminosity does not vary with galactic latitude in the interval covered by our sample. 

\begin{figure}
    \centering
      \includegraphics[angle=-90,width=0.5\textwidth,viewport=0 90 550 650]{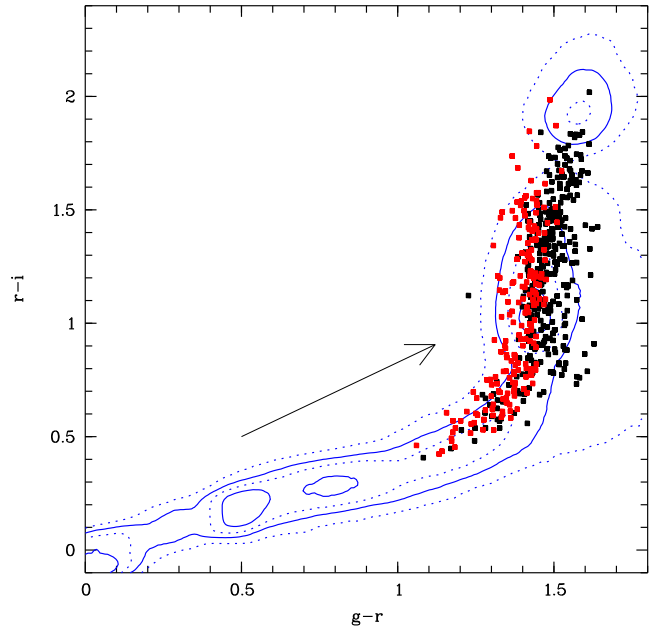}
    \caption{Distribution of X-ray active stars in the $g-r$ / $r-i$ diagram for two ranges of X-ray luminosities. Black dots Log(Lx)$<$29.3. Red crosses  Log(Lx)$>$29.3. The arrow shows the direction of interstellar reddening. Only sources with photometric errors below 0.1 mag in $g-r$ and $r-i$ are shown. Density contours show the position of all stars identified in the SEGUE programme.}
    \label{plotLowAndHighLxStars}
\end{figure}

We show in Fig. \ref{plotLowAndHighLxStars} the distribution of X-ray active stars in the $g-r$ / $r-i$ diagram for two ranges of X-ray luminosities. It can be readily seen that the locus of X-ray bright sources is shifted by $\sim$ 0.1 mag in colour above that occupied by low \Lx\ active coronae and by spectroscopically identified stars from SEGUE in general. Active stars appear to be bluer in $g-r$ for a given $r-i$. The effect is particularly clear for the reddest stars with $g-r$ $\ga$ 1.3, corresponding to M0 types and later. This shift cannot be due to the lack of reddening correction since the most X-ray luminous stars that are expected to be the most remote and absorbed ones should appear redder than the closest low \Lx\ stars, a trend opposite to what is observed. We also checked with the isochrones of \cite{girardi2004} designed for the SDSS band passes that age or metallicity effects were unable to explain the different colour/colour tracks of the low and high \Lx\ stars. Similarly, an enhanced \Halp\ emission expected to be especially important for late M stars ($g-r\ > 1.3$) can be excluded since it would yield larger $g-r$.  Interestingly, \cite{covey2008} report that the counterparts of their extended ChaMP stellar X-ray survey do exhibit a $\sim$ 0.1 bluer $u-g$ index than average low-mass stars, although they do not report any significant change in the $g-r$ colour index. Comparing with their work is however not straightforward. While the colour/colour track followed on average by our active coronae is consistent with that of \cite{covey2007} for the entire SDSS and therefore does not indeed contradicts the results of \cite{covey2008}, the difference we find rather arises among two groups of active stars.  The SDSS M stars templates compiled by \cite{bochanski2007} could indicate a similar trend in the $u-g$ colour index between active and inactive stars with however a large intrinsic scatter, while no such effect occurs in $g-r$. Unfortunately, we have too few good $u$ band measurements to be able to confirm the $u-g$ trend seen by \cite{covey2008}. We agree with these authors that low-level optical flaring might be responsible for the bluer colours seen in X-ray active M dwarfs. 

Active coronae emit essentially thin thermal spectra dominated by a series of narrow emission lines superposed on a weak continuum. In most cases, two thermal components are required to satisfactorily represent the observed energy distribution \citep[see a recent review in][]{guedel2009}. X-ray studies of open clusters and field stars of different ages led to a relatively coherent picture linking stellar rotation rates, overall X-ray luminosity and X-ray temperatures. Whereas the young (age $<$ 1\,Myr) stars in Orion exhibit X-ray spectra with $kT_1\,\sim$ 0.8\,keV and  $kT_2\,\sim$ 2.9\,keV, the analysis of $\sim$ 115\,Myr old Pleiades stars yields $kT_1\,\sim$ 0.4\,keV and  $kT_2\,\sim$ 1.1\,keV, while the X-ray corona of our Sun can also be characterised by 2T spectrum with $kT1\,\sim$ 0.2\,keV and  $kT2\,\sim$ 0.6\,keV \citep[see][and references therein]{sung2008}. Using ROSAT and ASCA observations of a dozen of carefully selected stars, \cite{guedel1997} established that the overall X-ray luminosity, the temperature of the two components, and the emission measurement ratio of the hot to the cool plasma were all decreasing with age and rotation rate. The range of X-ray luminosity observed in our survey suggests ages younger than $\sim$ 2\,Gyr. 
     
\begin{figure}
    \centering
    \includegraphics[angle=-90,width=0.45\textwidth,viewport=0 90 550 650]{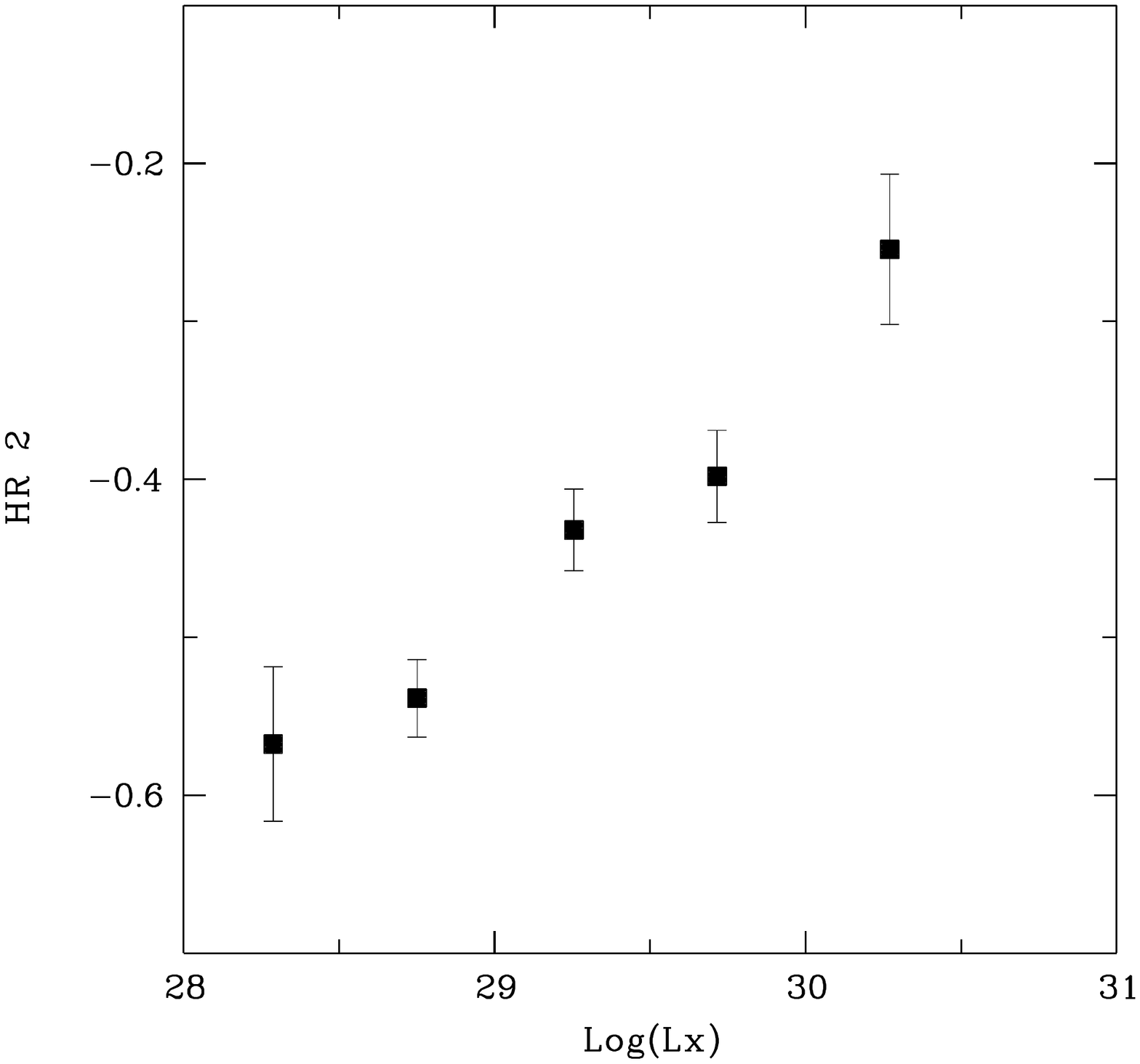}
    \includegraphics[angle=-90,width=0.45\textwidth,viewport=0 90 550 650]{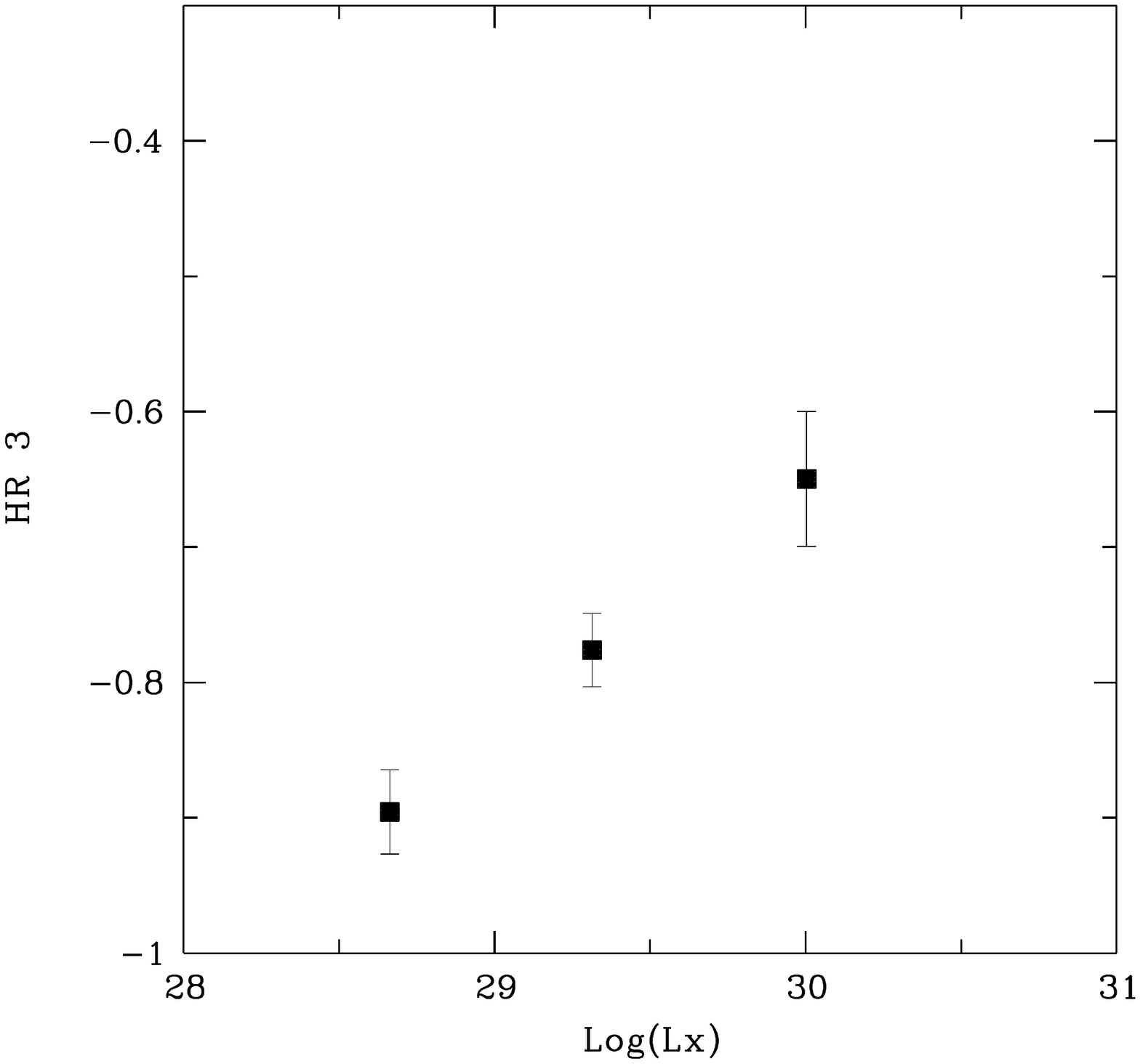}
    \caption{Variation of the EPIC pn Hardness ratios 2 and 3 with mean X-ray luminosity for stellar X-ray sources with HR errors below 0.2. We plot here the median and associated errors.}
    \label{plotHr2Lxcurve}
\end{figure}

We thus investigated whether the X-ray properties of our identified active coronae were depending on luminosity. For that purpose we selected the 296 (149) identified stellar X-ray sources having errors on 2XMMi hardness ratio 2 (3) of less than 0.2. This sample was then split into five ranges of luminosity for which the mean HR2 was computed, and in order to accommodate the lower number of sources in the harder band, into three bands of X-ray luminosities for HR3. Hardness ratio 2 and 3 measure the relative count rates in the energy ranges 0.5-1.0\,keV and 1.0-2.0\,keV and 1.0-2.0\,keV and 2.0-4.5\,keV respectively. They are therefore weakly dependent on the EPIC filter used and vary little within the range of \nh\ applicable to the present survey. HR2 and HR3 can thus be considered as indicators of the intrinsic shape of the X-ray energy distribution well suited to the range of temperatures exhibited by stellar coronae. We show in Fig. \ref{plotHr2Lxcurve} the variation of the median HR2 and HR3 with X-ray luminosity. A clear spectral hardening accompanies the luminosity increase. Assuming a single temperature plasma undergoing \nh $\sim$ 1.7 $\times$ 10$^{20}$ cm$^{-2}$ the HR2 value would imply a thin thermal temperature of $\sim$ 0.45 for the lowest \Lx\ bin and 0.72\,keV for the largest \Lx , while for the same range of X-ray luminosities the HR3 value would indicate temperatures from $\sim$ 0.45\,keV to $\geq$ 1.4\,keV. The existence of a second hotter thin thermal component naturally accounts for this discrepancy. Unfortunately, using only two hardness ratios, it is impossible to fit both $kT_1$, $kT_2$ and the ratio of the emission measurements of the hot and cool components, which is also expected to vary with X-ray luminosity \citep{guedel1997}. In spite of this shortcoming, our data which benefit from the large throughput of XMM-Newton and of its capability to obtain detailed spectral information for an unprecedented number of sources, confirm for the first time the spectral hardening with X-ray luminosity in large groups of mixed age field objects. 

\section{Summary}

We present the results of the cross-identification of the 2XMMi catalogue containing over 220,000 XMM-Newton EPIC serendipitous sources with the Data Release 7 of the Sloan Digital Sky Survey consisting of 357 million unique objects and over 1.6 million spectra. In order to ensure the best statistical control on the quality of the cross-correlation process, we use a likelihood ratio scheme only based on the probability of spatial coincidence of the X-ray source with the optical candidate. Using an original method that does not rely on heavy Monte Carlo simulations, we are able to compute true probabilities of identifications taking into account the varying fraction of X-ray sources expected to have a SDSS match as function of galactic latitude. We discuss the details of the statistical method used and present reliability and completeness curves for the entire set of matches. A total of 30,000 unique X-ray sources have a match in the DR7 with a probability of identification above 90\%. At this threshold, the reliability of the total sample is of 98\%, i.e. we expect only 2\% of spurious cross-identifications, while the completeness is 77\%, meaning that we miss about a quarter of the true matches, which therefore appear with individual probabilities of identifications below the 90\% threshold. 

One of the most ambitious task attributed to the Survey Science Centre of the XMM-Newton satellite is the statistical identification of all serendipitous X-ray sources discovered in the large field of view of the EPIC cameras. The 2XMMi/DR7 cross-identified sample offers an interesting opportunity to test various strategies and methods which would eventually led to the classification of the XMM catalogues. \cite{pineau2008} presented a first attempt to reduce the parameter space dimension using Principal Component Analysis tools adapted to data affected by rather large instrumental errors and taking into account the intrinsic scatter of the measured quantities. These authors also tested the relative merits of various classification methods.
We use the SDSS DR7 spectroscopic catalogue to build an identified sample made of objects of known nature to which we add a few entries extracted from specific additional catalogues (e.g. Cataclysmic Variables). We find that the most discriminating diagrams involve the $f_x/f_r$ flux ratio and various other parameters such as the $g-i$ colour index, EPIC hardness ratios, and the extent of the optical source. Active galactic nuclei more luminous than 10$^{42}$\ergs\ have $f_x/f_r$ ratios well above those of most galactic stars in their range of $g-i$. Galaxies of lower X-ray luminosity overlap the region of the diagram covered by active coronae, but can be easily separated from them on the basis of the optical extent. 

Not unexpectedly, active galactic nuclei constitute by far the most numerous class of X-ray emitters present in the cross-correlation. Optically unresolved UV excess quasars are the most frequent X-ray emitters encountered and cluster in a well defined area in the $(f_x/f_r)$ / $(g-i)$ diagram. The second largest group of X-ray detected AGN consists of optically extended objects, with similar or only slightly lower $f_x/f_r$ ratios, but exhibiting considerably redder $g-i$ colours. The very scarce SDSS spectroscopic coverage of this region does not allow constraining their true nature. However, relatively bright galaxies harbouring active nuclei with \Lx\ somewhat lower than those of UV excess quasars are likely to account for most of these objects. The vast majority of these "red" AGN exhibit X-ray hardness ratios consistent with those expected from a typical type I nucleus. However, a small fraction of these sources display evidence of enhanced photo-electric absorption that could be the signature of Compton Thin QSO2s, although other species of AGN can overlap the same parameter region. As \Lx\ (and $f_x/f_r$) decrease, extragalactic X-ray sources become spatially resolved in the optical and their $g-i$ colour indices tend to cluster around $g-i$ $\sim$ 1.3. In parallel, the hardness ratios of lower \Lx\ sources appear globally softer. We probably see here the effect of the increasing contribution of non-nuclear X-ray sources. Although some stars share $f_x/f_r$ ratios and $g-i$ colours similar to some galaxies, taking into account the extent of the optical object allows an almost perfect separation of the galactic and extragalactic source populations. As expected, the latest M stars display larger  $f_x/f_r$ ratios than earlier active coronae. 

Using a sample of reliable X-ray and optical Compton Thin and Compton Thick QSO2s we investigated the possibility of preferentially finding these objects in some specific range of X-ray and optical parameters. We find that Compton Thick QSO2 lack outstanding observational optical and X-ray photometric signatures that may be used to select sub-samples with greatly enhanced densities. However, the photo-electric absorption detectable in X-rays for relatively low-redshift AGN could be used to set apart groups of Compton Thin candidates with a reasonable success rate.   

We also find that X-ray bright stars appear slightly bluer in $g-r$ than less active coronae. The origin of this effect is not clear, but could be related to the UV and blue micro flaring occurring in many of the active late type K and M stars. Assuming that most active stars are indeed main sequence, we find a strong dependency of EPIC hardness ratios 2 and 3 with the inferred X-ray luminosity, the most luminous coronae being those emitting with the highest temperature. 

The cross-correlation of two large catalogues such as the 2XMMi and the DR7 of the SDSS paves the way to a wide range of investigations. In addition, the SDSS database contains line fluxes for over a million of spectroscopic targets, while the 2XMMi catalogue provides X-ray spectra and time series for the 44,000 brightest serendipitous XMM-Newton sources. These rich data sets offer unprecedented opportunities for studying how the micro-physics of the optically emitting regions relates to X-ray properties. Eventually, such studies, in particular those aiming at finding rare outliers, rely on a solid control of the cross-correlation and cross-identification statistics. 

We presented the details of an algorithm based on the classical likelihood ratio, well adapted to cross-correlation of large catalogues and providing useful probabilities of identification. This algorithm is now available as a plug-in for the Aladin VO portal\footnote{The plugin and its documentation can be downloaded from the home page of the XCat-DB (http://xcatdb.u-strasbg.fr/).}. 

\bibliography{XCorrelation2XMMiSDSS}

\begin{acknowledgements}
      We thank an anonymous referee for providing constructive comments and help in improving the contents of this paper.
       
      F.-X. Pineau acknowledges support from \emph{CNRS}, \emph{CNES} and from the \emph{Region Alsace}.     
      
      Funding for the SDSS and SDSS-II has been provided by the Alfred P. Sloan Foundation, the Participating Institutions, the National Science Foundation, the U.S. Department of Energy, the National Aeronautics and Space Administration, the Japanese Monbukagakusho, the Max Planck Society, and the Higher Education Funding Council for England. The SDSS Web Site is http://www.sdss.org/.

    The SDSS is managed by the Astrophysical Research Consortium for the Participating Institutions. The Participating Institutions are the American Museum of Natural History, Astrophysical Institute Potsdam, University of Basel, University of Cambridge, Case Western Reserve University, University of Chicago, Drexel University, Fermilab, the Institute for Advanced Study, the Japan Participation Group, Johns Hopkins University, the Joint Institute for Nuclear Astrophysics, the Kavli Institute for Particle Astrophysics and Cosmology, the Korean Scientist Group, the Chinese Academy of Sciences (LAMOST), Los Alamos National Laboratory, the Max-Planck-Institute for Astronomy (MPIA), the Max-Planck-Institute for Astrophysics (MPA), New Mexico State University, Ohio State University, University of Pittsburgh, University of Portsmouth, Princeton University, the United States Naval Observatory, and the University of Washington.
    
This research has made use of the SIMBAD database and of the VizieR catalogue access tool, operated at CDS, Strasbourg, France. 

\end{acknowledgements}

\begin{appendix} 

    \section{Transformation of error ellipses in the new frame}\label{apA}

        \subsection{Spherical trigonometry}

            On the unit sphere of centre $C$, we consider the spherical triangle $NXO$ where $N$ is the north pole, $X$ and $O$ are
            the positions of the X-ray and the optical sources respectively.
            We write
            \begin{itemize}
            \item $x$, $n$, $o$ the angular distances between $N$ and $O$, $O$ and $X$, $N$ and $X$ respectively.
            \item $\hat{N}$, $\hat{X}$ and $\hat{O}$ the dihedral angles between planes $NCX$ and $NCO$, $XCN$ and $XCO$, $OCN$ and $OCX$ respectively.
                    Those angles in the range $]-180\ ^{\circ},180\ ^{\circ}]$ and are define as positive in the East direction.
            \end{itemize}
            We immediately have $\hat{N} = \alpha_O-\alpha_X$, $x = 90\ ^{\circ}-\delta_O$, $o = 90\ ^{\circ}-\delta_X$
            and $n=d$ the angular distance between the two sources, which is given by the Haversine formula
            \begin{equation}
                    d = 2 \cdot \arcsin{\sqrt{\sin^2 \frac{\delta_o - \delta_X}{2} +\sin^2{\frac{\alpha_o-\alpha_X}{2}}  \cos{\delta_X} \cos{\delta_o}}} \ .
                    \label{eq:angular_distance}
                \end{equation}
    
            The spherical trigonometry gives the formula
            \begin{equation}
                \frac{\sin{\hat{N}}}{n}=\frac{\sin{\hat{X}}}{x}=\frac{\sin{\hat{O}}}{o} \ .
            \end{equation}
            As $n$, $x$ and $o$ are positive, we notice that $\hat{N}$, $\hat{X}$ and $\hat{O}$ have the same sign, 
            which depends on $\alpha_O - \alpha_X$.
            As the function $arcsin$ returns values $\in [-90,+90]$, we deduce two possible values for both angle $\hat{X}$ and $\hat{O}$.
            The cosine of half-angles $\hat{X}/2$ and $\hat{O}/2$ enable to know if $|\hat{X}|>90\ ^{\circ}$ and if $|\hat{O}|>90\ ^{\circ}$ respectively.
            From the spherical trigonometry formulae
            \begin{eqnarray}
                s                       & = & \frac{1}{2}(n+x+o) \\
                \cos{\frac{\hat{X}}{2}} & = & \sqrt{\frac{\sin{s}\sin{(s-x)}}{\sin{n}\sin{o}}}\\
                \cos{\frac{\hat{O}}{2}} & = & \sqrt{\frac{\sin{s}\sin{(s-o)}}{\sin{n}\sin{x}}}
            \end{eqnarray}
            We derive the values of angles $\hat{X}$ and $\hat{O}$:
            \begin{equation}
                \hat{X} = \left\{ 
                    \begin{array}{lll}
                        180 - \arcsin{\frac{x\cdot \sin{\hat{N}}}{n}} & \mathrm{ if }&\\
		        \ \ \ \ \ \ \cos{\frac{\hat{X}}{2}} > \frac{\sqrt{2}}{2} & \mathrm{ and } & \arcsin{\frac{x\cdot \sin{\hat{N}}}{n}} \geq 0 \\
                        -180 - \arcsin{\frac{x\cdot \sin{\hat{N}}}{n}} & \mathrm{ if }&\\
			\ \ \ \ \ \ \cos{\frac{\hat{X}}{2}} > \frac{\sqrt{2}}{2} & \mathrm{ and } & \arcsin{\frac{x\cdot \sin{\hat{N}}}{n}}  <  0 \\ 
                        \arcsin{\frac{x\cdot \sin{\hat{N}}}{n}}      & \mathrm{ else} &\\
                    \end{array}
                \right.
            \end{equation}
            \begin{equation}
                \hat{O} = \left\{ 
                    \begin{array}{lll}
                        180 - \arcsin{\frac{o\cdot \sin{\hat{N}}}{n}} & \mathrm{ if }&\\
			\ \ \ \ \ \ \cos{\frac{\hat{O}}{2}} > \frac{\sqrt{2}}{2} & \mathrm{ and } & \arcsin{\frac{o\cdot \sin{\hat{N}}}{n}} \geq 0 \\
                        -180 - \arcsin{\frac{o\cdot \sin{\hat{N}}}{n}} & \mathrm{ if }&\\
			\ \ \ \ \ \ \cos{\frac{\hat{O}}{2}} > \frac{\sqrt{2}}{2} & \mathrm{ and } & \arcsin{\frac{o\cdot \sin{\hat{N}}}{n}}  <  0 \\ 
                        \arcsin{\frac{o\cdot \sin{\hat{N}}}{n}}       & \mathrm{ else} & \\ 
                    \end{array}
                \right.
            \end{equation}

        \subsection{Transformation of an error ellipse into a variance-covariance matrix}

            Our starting frame $F_s$ is the plane centred in $X$ and perpendicular in $X$ to $(CX)$. It has as its $x$-axis the East direction and as its $y$-axis the North direction. Errors on positions in astronomy are almost always given in the form of an ellipse
of major axis $a$, minor axis $b$ and of angle $\theta$ between the north pole and the major axis part oriented in the East direction.
            So $\theta \in [0,180[$.
            We can transform the ellipse into a variance-covariance matrix in the frame $F_s$. 
            The frame in which the ellipse is in its canonical form is obtain by rotating the frame $F_s$ of an angle $\beta=90-\theta$.
            Applying a rotation of angle $-\beta$ on the canonical variance-covariance matrix 
            give us the expression of the variance-covariance matrix in $F_s$: 
                \begin{equation}
                    V_{F_s} = 
                    \Big (
                        \begin{array}{cc}
                            \cos\beta & \sin\beta\\
                           -\sin\beta & \cos\beta\\
                        \end{array}
                    \Big )
                    \Big (
                        \begin{array}{cc}
                            a & 0\\
                            0 & b\\
                        \end{array}
                    \Big )
                    \Big (
                        \begin{array}{cc}
                           \cos\beta & -\sin\beta\\
                           \sin\beta & \cos\beta\\
                        \end{array}
                    \Big ) \ .
                \end{equation}
            It leads to
            \begin{eqnarray}
                \sigma_x^2  & = & a\sin^2\theta+b\cos^2\theta \\
                \sigma_y^2 & = & a\cos^2\theta+b\sin^2\theta \\
                \rho\sigma_x\sigma_y & = & \cos\theta\sin\theta(a-b)   \ . 
            \end{eqnarray}

       \subsection{Variance-covariance matrices in our new frame}

            As mention in Sect. \ref{sec:optcandselect}, our new frame $F_n$ is centred in $X$, has as its $x$-axis the direction of the perpendicular in $X$ of $(CO)$ in the plane $XCO$ oriented in the East part and as its $y$-axis the perpendicular in $X$ of the plane $XCO$ oriented in the North part.
            Given the results of the previous section, the new frame is obtained from the starting one ($F_s$) by a rotation of angle 
            $\Omega=90-\hat{X}$ if $\hat{X}$ is positive and $\Omega=-90-\hat{X}$ if $\hat{X}$ is negative.
            The variance-covariance matrix of the errors on positions can be expressed in the new frame $F_n$ by a matrix $V_n$
            \begin{equation}
                    V_n = 
                    \Big (
                        \begin{array}{cc}
                            \cos\Omega & \sin\Omega\\
                           -\sin\Omega & \cos\Omega\\
                        \end{array}
                    \Big )
                    \Big (
                        \begin{array}{cc}
                            \sigma_x^2 & \rho\sigma_x\sigma_y0\\
                            \rho\sigma_x\sigma_y & \sigma_y^2\\
                        \end{array}
                    \Big )
                    \Big (
                        \begin{array}{cc}
                           \cos\Omega & -\sin\Omega\\
                           \sin\Omega & \cos\Omega\\
                        \end{array}
                    \Big ) \ ,
            \end{equation}
            leading to
            \begin{eqnarray}
                \sigma_{x_n}^2  & = & \sigma_x^2\cos^2{\Omega} + \sigma_y^2\sin^2{\Omega} + 2\cos{\Omega}\sin{\Omega}\rho\sigma_x\sigma_y \\
                \sigma_{y_n}^2 & = & \sigma_x^2\sin^2{\Omega} + \sigma_y^2\cos^2{\Omega} - 2\cos{\Omega}\sin{\Omega}\rho\sigma_x\sigma_y \\
                \rho_n\sigma_{x_n}\sigma_{y_n} & = & \cos{\Omega}\sin{\Omega}(\sigma_y^2-\sigma_x^2)+(\cos^2{\Omega}-\sin^2{\Omega})\rho\sigma_x\sigma_y \ .
            \end{eqnarray}

       \subsection{Convolution product of two independent 2D Gaussians}

            The errors on the X-ray and the optical source are two Gaussian random variables $Z_X$ and $Z_o$
            with variance-covariance matrices
            \begin{equation}
                    V_{Z_X}=
                    \Big (
                        \begin{array}{cc}
                            \sigma^2_{x_X} & \rho_{X}\sigma_{x_X}\sigma_{y_X}\\
                            \rho_{X}\sigma_{x_X}\sigma_{y_X} & \sigma^2_{y_X}\\
                        \end{array}
                    \Big )
                    \mathrm{, }
                    V_{Z_o}=
                    \Big (
                        \begin{array}{cc}
                            \sigma^2_{x_o} & \rho_{o}\sigma_{x_o}\sigma_{y_o}\\
                            \rho_{o}\sigma_{x_o}\sigma_{y_o} & \sigma^2_{y_o}\\
                        \end{array}
                    \Big ) \ .
            \end{equation}
            The random variable $Z=Z_x+Z_o$ is define by the convolution product of $Z_X$ by $Z_o$.
            We know that the sum of 2 Gaussians is a Gaussian with a variance-covariance matrix:
            \begin{equation}
                    V_{Z}=
                    \Big (
                        \begin{array}{cc}
                            \sigma^2_{x_X+x_o} & \rho\sigma_{x_X+x_o}\sigma_{y_X+y_o}\\
                            \rho\sigma_{x_X+x_o}\sigma_{y_X+y_o} & \sigma^2_{y_X+y_o}\\
                        \end{array}
                    \Big ) \ .
             \end{equation}
            As $Z_X$ and $Z_o$ are independent, $x_X$ and $x_o$ are independent. Idem for $y_X$ and $y_o$.
            We thus have $Cov(x_X,x_o)=Cov(y_X,y_o)=0$.
            The variance formula $Var(a+b)=Var(a)+Var(b)+2Cov(a,b)$ leads to
            \begin{eqnarray}
                \sigma^2_{x_X+x_o} & = & \sigma^2_{x_X} + \sigma^2_{x_X} \\
                \sigma^2_{y_X+y_o} & = & \sigma^2_{y_X} + \sigma^2_{y_X}
            \end{eqnarray} \ .
            The covariance is given by $\rho\sigma_{x_X+x_o}\sigma_{y_X+y_o} = E\{(x_X+x_o)(y_X+y_o)\}-E\{(x_X+x_o)\}E\{(y_X+y_o)\}$.
            But as our four 1D Gaussian $x_X$, $x_o$, $y_X$ and $y_o$ are centred, their mean is null and thus
            $E\{(x_X+x_o)\}=E\{x_X\}+E\{x_o\}=0$ and $E\{(y_X+y_o)\}=E\{y_X\}+E\{y_o\}=0$.
            It leads to $\rho\sigma_{x_X+x_o}\sigma_{y_X+y_o}=E\{(x_Xy_X)\}+E\{(y_Xy_o)\}+E\{(x_Xy_o)\}+E\{(y_Xx_o)\}$, which is a sum
            of covariance since our four distributions are centred. We already mentioned that the covariance of $x_X$ and $x_o$ is null.
            Idem for $y_X$ and $y_o$. We finally obtain
            \begin{equation}
                \rho\sigma_{x_X+x_o}\sigma_{y_X+y_o} = \rho_{X}\sigma_{x_X}\sigma_{y_X} + \rho_{o}\sigma_{x_o}\sigma_{y_o}
            \end{equation}

\end{appendix}

\begin{appendix}

\section{Estimate of the local density} \label{part:density_estimation}
            
In the $1 \sigma$ ellipse (Eq. \ref{eq:r}), which is the elementary surface, the Poissonian density of sources is $\lambda = \pi\sigma_M\sigma_m f(\alpha_x,\delta_x,m_o)$. The accuracy of the likelihood ratio (Eq: \ref{eq:lr}) depends sensitively on the local density estimation $f(\alpha_x,\delta_x,m_o)$. A simple manner to estimate the local density of the SDSS-DR7 catalogue is to use a k nearest neighbour ({\em knn}) averaging. In this case, the estimated densities have a well known error: $\sqrt{k}/\pi d_k^2$, with $d_k$ the distance of the $k^{th}$ neighbour from the X-ray source. However, {\em knn} averaging has the impeding property of being non-differentiable. In fact, two candidates with similar magnitudes -- or two nearby sources with the same magnitude -- could have quite different $LR$s. The Voronoi tessellation and the use of wavelets also seems too complicated and time-consuming for our simple purpose.

Considering these drawbacks, we preferred to use the sample-point density estimator, which is a kernel smoothing, using the Epanechnikov profile as kernel function and the distance to the 100$^{th}$ nearest neighbours as bandwidth.

Kernel smoothing -- also called Parzen window technique or kernel density estimation --
can be performed with a fixed or a variable bandwidth.
\citet{DBLP:journals/csda/ZhangKH06} provide Markov Chain Monte Carlo (MCMC) algorithms for estimating
optimal data-driven fixed bandwidths. These algorithms involve the creation of a density map.
In our case we consider for each candidate the density of sources at least as bright as the candidate.
We therefore should compute roughly as many density maps as the number of candidates and apply several times
the algorithms. Since the MCMC technique in our case is time-consuming, it has been discarded.

We therefore investigated variable bandwidth kernel smoothing techniques.
The simplest one is the \emph{balloon} estimator \citep[see][]{603891}.
Unfortunately it usually fails to integrate to one.
We therefore preferred a \emph{sample-point density estimator} in which the estimate of the density at a
point $\vec{x}$ of a FOV containing $N$ sources at positions $\vec{x_i}\mbox{, }i\in[[1,N]]$ is given by
\begin{equation}
    \hat f(\vec{x}) = \sum\limits^N_{i=1}\frac{1}{h(\vec{x_i})^2} K_E\left(  \frac{||\vec{x}-\vec{x_i}||}{h(\vec{x_i})} \right) \ ,
\end{equation}
where $K$ is a kernel function and $h(\vec{x_i})$ a bandwidth depending on the position.
According to \citet{937550}, this estimator \emph{``is proved to be almost all the time much better than the fixed bandwidth estimator''}.
The article gives an usual form of the variable bandwidth $h(\vec{x_i})$:
\begin{equation}
    h(\vec{x_i}) = h_o\left[ \frac{\lambda}{f(\vec{x_i})} \right]^{1/2} \ .
\end{equation}
where $h_o$ represents a fixed bandwidth, $\lambda$ a proportionality constant impacting on the smoothness and
$f$ a pilot function. It is also mentioned that \emph{``the method is insensitive to the fine detail of the pilot estimate''},
and \emph{``a good initial choice is to take $\lambda$ as the geometric mean''} of the pilot function.\\

The selected pilot function stems from a {\em knn} averaging: $f(\vec{x_i})=\frac{k}{\pi d_k^2}$ where $d_k$ is the distance to the $k^{th}$ nearest neighbour at least as bright as the candidate. It is fast to compute and the discontinuities can be considered as negligible \emph{fine details}.
As suggested in \cite{937550}, we define $\lambda$ as the mean value of the pilot function: $\lambda=\frac{N}{\pi R_{fov}^2}$, where $N$ is the total number of sources in the field of view with a magnitude lower than or equal to that of the candidate, and where $R_{fov}$ is the radius of the FOV.
We used a fixed bandwidth, which corresponds to the area in which we find, on the average, $k$ sources: $h_o=\sqrt{\frac{k}{N}}R_{fov}$.
This leads to $h(\vec{x_i}) = d_k$. In order to complete our estimate of the local density we only need to define parameter $k$, the number of nearest neighbours being at least as bright as the candidate considered.

The smaller is $k$, the more local is the density, but the larger is the error on the density estimation. Using $k=100$ leads to a relative error on $LR$ of about 15\%, which corresponds to $6.5\%$ for $\log_{10}LR$. In addition, we implemented an algorithm handling border effects. 
	    
The selected kernel profile is the 2D Epanechnikov profile $K_E$:
\begin{equation}
    K_E(\tau) = \left\{ 
    \begin{array}{lll}
	\frac{2}{\pi}(1-\tau^2) & \mbox{if } & \tau < 1 \\
	0 & \mbox{else} & 
    \end{array} \ .
    \right.
\end{equation}
This truncated parabola allows computational optimisations thanks to its finite extent. Moreover, since it minimises the mean integrated square error \citep{513076}, this kernel maximises the quality of the density reconstruction.

\end{appendix}

\begin{appendix} 

    \section{Some details on the likelihood ratio}\label{LRBayesian}

        We consider the case where an X-ray source has only one candidate and formulate two hypotheses:
        \begin{description}
            \item[$H_{cp}$:] the candidate is the counterpart, it is not random.
            \item[$H_{spur}$:] the candidate is a random source belonging to a Poissonian distribution of sources of density $\lambda$.
        \end{description}
        The union of these two hypotheses gives the set of all possibilities: $P(H_{cp})+P(H_{spur}) = 1$, .i.e $H_{cp} = \bar{H}_{spur}$.
        The probability of finding a candidate at a distance $r$ of the X-ray source can therefore be written as:
        \begin{equation}
            P(r) = P(r|H_{cp})\cdot P(H_{cp})+P(r|H_{spur})\cdot P(H_{spur}) \ .
            \label{eq:pr}
        \end{equation}
        From conditional probabilities
        \begin{eqnarray}
            P(H_{cp}|r) & = & \frac{P(H_{cp}\cap r)}{P(r)}      \\
            P(r|H_{cp}) & = & \frac{P(H_{cp}\cap r)}{P(H_{cp})}
        \end{eqnarray}
        we deduce
        \begin{equation}
            P(H_{cp}|r) = \frac{P(r|H_{cp})\cdot P(H_{cp}) }{P(r)} \ ,
        \end{equation}
        which can be written (Eq. (\ref{eq:pr})) as:
        \begin{equation}
            P(H_{cp}|r) = \frac{P(r| H_{cp})\cdot P(H_{cp}) }{P(r| H_{cp})\cdot P(H_{cp})+P(r| H_{spur})\cdot P(H_{spur})} \ .
        \end{equation}
        We transform that equation into:
        \begin{equation}
            P(H_{cp}|r) = \frac{1}{1+\frac{1}{\frac{P(H_{cp})}{P(H_{spur})}LR}} \ ,
            \label{eq:phcp}
        \end{equation}
        to exhibit the likelihood ratio
        \begin{equation}
            LR(r) = \frac{P(r| H_{cp})}{P(r| H_{spur})} \ .
        \end{equation}

        The remaining term $\frac{P(H_{cp})}{P(H_{spur})}$ has to be estimated to compute $P(H_{cp}|r)$, the probability that the candidate
        is a real counterpart.
        Considering the estimate of the number of real counterparts divided by the number of spurious associations, it leads to the $\theta/(1-\theta)$ factor of \citet{1977A&AS...28..211D}.

        If we add a prior knowledge on the magnitude, the formula (\ref{eq:phcp}) is unchanged except that it now expresses $P(H_{cp}|r\cap m)$ and the $LR$ becomes
        \begin{equation}
            LR(r\cap m) = \frac{P(r\cap m| H_{cp})}{P(r\cap m| H_{spur})} = \frac{P(r| H_{cp})P(m| H_{cp})}{P(r| H_{spur})P(m| H_{spur})} \ .
            \label{eq:lrextended}
        \end{equation}
        The term $P(r\cap m| H_{spur})=P(r| H_{spur})P(m| H_{cp})$ expresses the probability of having a spurious source of magnitude $m$.
        It is the Poissonian local distribution of sources of magnitude $m$.
        The term $P(r| H_{cp})P(m| H_{cp})$ is the classical 2D Gaussian distribution times the probability of having, among all real counterparts,
        a counterpart of magnitude $m$. It can be seen as the distribution of counterpart sources according to their magnitudes and
        correspond to the $q(m)$ factor in \citet{2007ApJS..172..353B} and of $q(m,c)$ in \citet{1992MNRAS.259..413S} if we ignore object types.

\end{appendix}

\begin{appendix}

\section{Computation of the spurious $LR$ distribution}\label{spurLR}

Let us first consider the case of one X-ray and one optical source of magnitude $m$ in a FOV. The probability of observing the random X-ray source at a given point $(\alpha_X,\delta_X)$ of the FOV is given by the distribution $g_X(\alpha_X,\delta_X)$ of the X-ray sources in the FOV. The chance for the optical source of magnitude $m$ to be randomly associated with the X-ray source is given by the integral of the distribution of optical sources at least as bright as $m$ over the surface of the convolution ellipse $S_{x,o}=k_\gamma^2\pi\sigma_M\sigma_m$. The probability that the association has a $LR$ in a bin $\Delta LR$ is then given by the integral of the distribution of optical sources at least as bright as $m$ over the surface the bin occupies inside the convolution ellipse $S_{x,o}$: $S_{x,o}(\Delta LR)$. Finally, the sum of these probabilities for all X-ray and archival sources provides an estimate of the number of spurious associations by $LR$ bin. Considering now the case of $N$ X-ray and $M$ optical sources in the FOV leads to the formula of $N_{spur}(\Delta LR)$
   \begin{equation}
   \sum\limits_{x=1}^N \iint\limits_{S_{FOV}} g_X(\alpha_X,\delta_X) \left( \sum\limits_{o=1}^M \iint\limits_{S_{x,o}(\Delta LR)} g_o(\alpha_o,\delta_o,m)   \mathrm{d}\alpha_o \mathrm{d}\delta_o \right) \mathrm{d}\alpha_X\mathrm{d}\delta_x \ ,
   \label{eq:nqpur1}
   \end{equation}
where $S_{x,o}(\Delta LR)$ depends on the X-ray source, the optical source, and the position of the X-ray source and can therefore be written as $S_{x,o}(\Delta LR,X,o,\alpha_X,\delta_X)$.

A possible manner to make the problem simpler is to assume that $g_o(\alpha_o,\delta_o,m)$ is roughly constant inside the surface $S_{x,o}(\Delta LR)$. Eq. (\ref{eq:nqpur1}) can then be written as
\begin{equation}
     \iint\limits_{S_{FOV}} g_X(\alpha_X,\delta_X)
                            \left(  \sum\limits_{x=1}^N\sum\limits_{o=1}^M S_{x,o}(\Delta LR)g_o(\alpha_X,\delta_X,m) \ .
                            \right) \mathrm{d}\alpha_X\mathrm{d}\delta_x
\end{equation}
In order to estimate $N_{spur}(\Delta LR)$ we have to compute a complete distribution map of the X-ray sources and as many distribution maps as the number of distinct optical source magnitudes. Since such a computation is potentially time-consuming, we may suppose that optical sources are uniformly distributed in the FOV and have as local density the mean value of their local densities. This less rigorous approach leads to the simple formula
\begin{equation}
    N_{spur}(\Delta LR) = \sum\limits_{x=1}^N\sum\limits_{o=1}^M S_{x,o}(\Delta LR)\left/S_{FOV} \right. \ .
    \label{eq:nspur}
\end{equation}

The quantity $S_{x,o}(\Delta LR)$ can be computed as follows: the radius corresponding to a given $LR$ is easily derived from Eq. (\ref{eq:lr}):
\begin{equation}
    r^2(LR) = \left\{ 
    \begin{array}{lll}
    	0			      & \mbox{if } & LR \geq LR_{max} = \frac{1}{2\lambda} \\
    	\ln \frac{1}{(2\lambda LR)^2} & \mbox{if } & LR_{min} < LR < LR_{max}\\
    	k_\gamma^2		      & \mbox{if } & LR \leq LR_{min} = \frac{1}{2\lambda}e^{\frac{1}{2}k_\gamma^2} 
    \end{array}
    \right.
\end{equation}
Thus for a bin $\Delta LR = LR_2 - LR_1 $, $LR_2 > LR_1$, if we have $LR_1 \leq LR_{max}$ and $LR_2 \geq LR_{min}$, the surface of the bin of $LR$ in the ellipse is expressed by
\begin{equation}
    S_{x,o}(\Delta LR) = 2\pi\sigma_M\sigma_m\ln \frac{min(max(LR_2,LR_{min}),LR_{max})}{max(min(LR_{max},LR_1),LR_{min})} \ .
\end{equation}

In principle, we should remove the optical sources which are the real counterparts since they are not randomly distributed. However, their fraction can be neglected at low $LR$, and at very high $LR$ their influence on the final probability of identification is insignificant. For real data, the source distribution is not completely Poissonian because two sources cannot be infinitely close due to the PSF, the resolution of the instrument, and the extraction algorithm. There is therefore a small surface around each counterpart which cannot contain another source. Neglecting this can lead to slightly overestimate the rate of spurious associations. 
\end{appendix}                 

\begin{appendix}

\section{Implementation}\label{implementation}
All software was written in Java. 2XMMi sources were retrieved using direct SQL queries -- through a Java DataBase Connection -- on the XCat-DB. SDSS DR7 sources were collected from the VO thanks to the ConeSearch protocol, the correlation has thus been performed independently for each XMM observation. 

Local density estimations required a lot of $k$-nearest-neighbours search. Moreover, for each candidate we are only interested in sources with a magnitude smaller than or equal to that of the candidate. In order to quickly perform such {\em knn} queries, SDSS DR7 sources have been stored in quadtrees especially designed for that purpose: each leaf of the tree stores the greatest magnitude among those of the sources held in its sub-tree ; {\em knn} queries contain an extra parameter which is the magnitude of the candidate; in addition to the distance criterion, we thus easily add a magnitude criterion to stop the search in a sub-tree.

A quadtree scheme was preferred to a kdtree because it is fastest to build and easier to update. It is then possible to build the quadtree during the parsing of the VOTable returned by the ConeSearch query.

\end{appendix}

\end{document}